\documentclass{article}

\usepackage{style}

\begin{document}

\setcounter{totalnumber}{10}

\title{PDF modeling of near-wall turbulent flows:\\
A New model, Weak second-order scheme and a numerical study in a Hybrid 
configuration }
\author{Sergio Chibbaro $^{1}$ and Jean-Pierre Minier $^{2}$ \vspace{2mm} \\
$^{1}$ Institut Jean Le Rond D'Alembert University Paris 6, \\
4, place jussieu 75252 Paris Cedex 05\\
sergio.chibbaro@upmc.fr \\
$^{2}$ Electricit\'e de France, Div. R\&D, MFTT, \\
6 Quai Watier, 78400 Chatou, France \vspace{2mm}}
\maketitle

\begin{abstract}
In this work, we discuss some points relevant for stochastic modelling of one- and two-phase turbulent flows. In the framework of stochastic modelling, also referred to PDF approach, we propose a new Langevin model including all viscosity effects and thus that is consistent with viscous Navier-Stokes equations. In the second part of the work, we show how to develop a second-order unconditionally stable numerical scheme for the stochastic equations proposed. Accuracy and consistency of the numerical scheme is demonstrated analytically. In the last part of the work, we study the fluid flow in a channel flow with the proposed viscous method. A peculiar approach is chosen: the flow is solved with a Eulerian method and after with the Lagrangian model proposed which uses some of the Eulerian quantities. In this way attention is devoted to the issue of consistency in hybrid Eulerian/Lagrangian methods. It is shown that the coupling is important indeed and that to couple the Lagrangian model to an Eulerian one which is not consistent with the same turbulence physics leads to large errors. This part of the work complements a recent article [Chibbaro and Minier {\it International Journal of Multiphase flows} submitted (arXiv:0912.2045)]. 
\end{abstract}

\section{Introduction}

Stochastic modeling approaches 
(also referred to as Probability Density Function (PDF) methods),
once developed in statistical mechanics
and solid state physics \cite{Ott_96},
have been shown to be very powerful 
for the study of turbulent fluids in presence of complex physics, 
in particular, for turbulent combustion flows \cite{Pop_85}
and for polydispersed two-phase flows \cite{Min_01}.
The main advantage of the PDF approach over 
conventional moment-closure methods is its ability
to reproduce convection and non-linear source terms
without approximations. 
Moreover, the information available by using such methods
is the complete PDF (though modeled) and, thus, 
all the statistical moments which are related.

In PDF methods, turbulent closure is achieved through
a modeled transport equation for the joint PDF of some chosen variables,
which constitute the {\it state vector} of the process.
In this work (and almost ever) the one-particle 
Lagrangian PDF is considered, 
from which the one-point one-time joint Eulerian PDF
can be extracted.
The resulting modeled one-particle PDF transport equation
is a {\it Fokker-Planck} equation \cite{Ott_96},
which is a high dimensional scalar equation.
Thus, traditional numerical techniques such as finite-volume 
and finite-difference methods are possible, in theory,
but are not suitable to solve PDF equations in practice, 
since the computational cost increases exponentionally with the number
of dimensions.
On the other hand, Fokker-Planck equations are equivalent (in a weak sense) 
to a set of stochastic differential equations (SDEs). 
In this case, the PDF is represented by an ensemble of 
Lagrangian stochastic particles
whose properties are driven by the model SDEs.
These stochastic particles can be regarded as samples of the PDF
and following these particles in time
represents a dynamical Monte Carlo method,
whose computational cost increases only linearly 
with the number of particles.
For this reason, the Monte Carlo method (or particle stochastic approach)
is usually
chosen to solve high dimensional equations and,
in particular,  PDF equations.

In turbulent two-phase flows, the SDEs equations of the model
contain several mean fields
and they have the general form
\begin{equation} \label{eq:MK}
dZ_{i}(t) =
A_{i}(t,{\bf Z},\lra{f({\bf Z})})\,dt+
\sum_j B_{ij}(t,{\bf Z},\lra{f({\bf Z})})\,dW_j(t),
\end{equation}
where the
$\lra{\;}$ operator stands for the mathematical expectation.
One way to compute mean fields is to extract them directly
from the particle properties with the help of a mesh,
this is the stand-alone method.
This method is fully consistent but  suffers, nevertheless,
from some drawbacks mainly due to the statistical fluctuations
in the particle mean fields\cite{Xu_99}.
On the other hand, hybrid methods,
where the mean fields are provided by a coupled different numerical method,
are also possible.
These hybrid methods are based
on particle-mesh techniques \cite{Hoc_88}: the mean-field
equations are solved on a mesh by standard discretization techniques
whereas the dynamics of the particles are still obtained 
by the time integration 
of the stochastic differential equations.
The main aim of such hybrid methods  is to improve 
the efficiency of numerical simulations without any important loss
of accuracy.
Indeed, those methods try to  conjugate 
the advantages of moment approach, which 
provides mean-fields free of statistical errors and
low computational costs, 
with those of PDF one, 
which are the accurate treatment of some non-linear phenomena
and the more detailed level of information. 

Recently, the possibility
to couple numerical methods of different kind has emerged
in several scientific domains,
for example the kinetic approach with the fluid one in 
DSMC simulations \cite{Bir_94}.
In fluid mechanics, different strategies have been explored 
to couple PDF methods with other approaches,
in particular, Moments/PDF \cite{Mur_01,Pei_06},
PDF/SPH (smooth particle hydrodynamics)\cite{Wel_98}
and LES/PDF \cite{Gic_02}.
In this work, we are dealing with two-phase flows
and  we choose the framework
of hybrid Moments/PDF approach, 
where, on the one hand, the fluid flow is approached 
by classical Eulerian moment method and, on the other hand,
the dynamics of the solid particles is directly simulated
by a stochastic process of the form (\ref{eq:MK}).
As mentioned above, the mean fields present in the stochastic model 
are provided by the Eulerian solver.
The present work complements a recent article~\cite{Chi_09ijmf}
and thus considers the same configuration treated there that is 
we limit ourselves to a fluid-fluid configuration,
that is, we are considering  tracer particles 
with zero diameter and, as a consequence, zero inertia.
This represents an asymptotic limit case
of the general two-phase flow configuration,
but numerically it preserves the same characteristics concerning the exchange
of variables between the different methods and, thus, it gives
a privileged position to 
address issues about the consistency of the method, 
which is a major point for such approaches and which can be investigated
with difficulty in more complex  situations.
Indeed, in  the Eulerian solver, classical second-order turbulent models
are generally used, 
while for the solution of SDEs different techniques must be used,
as explained later.
In this framework, several mean fields can be computed as duplicate fields
in the FV and particle algorithm,
which raises questions of consistency.
From a numerical point of view, we have chosen  
a turbulent channel flow as test-case
which represents a classical engineering 
situation.

Three main purposes characterize the present work.
The first one is to propose a new viscous PDF model  which
takes into account 
the viscous terms that are present in Navier-Stokes equations.
This kind of model may be useful when simulations of
low-Reynolds number or wall-bounded flows are taken under consideration,
since in those cases viscous terms are normally not negligible.
The viscous model proposed is shown to be consistent 
with exact first and second-order moment equations for velocity.
The second purpose is to propose a general and rigorous framework 
in which to develop numerical schemes for SDEs of the very general form
(\ref{eq:MK}). 
In particular, we develop an efficient, stable
and accurate numerical schemes for the integration of the trajectories of
the stochastic process.
This implies two main difficulties.
\begin{enumerate} 
\item[(a)] The first one arises from the nature of the stochastic models.
SDEs do not obey the rules of classical differential calculus and
one has to rely on the theory of stochastic processes
\cite{Kle_99}. In the present paper, It\^o's calculus is adopted and
therefore all SDEs are written in the \textit{It\^o sense}. For
stochastic processes, several convergence modes are possible
only weak convergence is under consideration,
since the purpose is just to estimate mean quantities. 
A discrete approximation ${\bf Y}_T$ ($T$ stands for a given stopping
time) converges in the weak sense with order $p$, if for any polynomial
$g$, there exists a constant $C$, function of $T$, such that
\begin{equation} 
|\lra{g({\bf Z}_T)}-\lra{g({\bf Y}_T)}| \leqslant C(T)\,\Delta t^{\,p}.
\end{equation}
Due to the mathematical definition of It\^o's integral,
numerical schemes developed for PDEs cannot be applied directly.
\item[(b)] The second difficulty is related to physical constraints.
As suggested in Minier \cite{Min_00},
the general stochastic model used to simulate general turbulent flows 
contains several characteristic time-scales.
When some of these time-scales become negligible (the system of SDEs is 
then stiff), various sub-systems of stochastic differential equations can be 
extracted. In other words, simplified stochastic models can be obtained
 from the general one. Our second objective is to put forward numerical schemes
 that can be still applicable, and that remain accurate, when the different 
time-scales go to zero~\cite{Min_03,Pei_06}.
Thus, it should be evident that this corresponds to a practical concern,
for in the numerical simulation of a complex flow, the time-scales may be
negligible in some areas of the flows. We want nevertheless the general 
numerical scheme to reproduce the correct physical behavior in these areas
with the same numerical efficiency. 
To overlook this point can cause severe problems and even flawed results.
For instance, in many studies, wall-bounded flows are considered where SDEs become stiff near the wall.
In these cases to use standard numerical schemes like standard euler is not stable.
The consequence is that very small time-steps have to be used to stabilise the numerical scheme.
That increases much the computational cost, on one hand, and makes results questionable, on the other hand.
\end{enumerate}
The third purpose of this work is to 
investigate numerically the behavior of the method,
even in the asymptotic limits, with the main attention devoted to
the issue of  consistency.
Indeed, hybrid methods are very attractive for simulations of
reactive and multiphase flows,
but the influence of the exchange of variables remains
to be completely understood.
It is emphasized here that it is possible to assess
the global consistency of the method,
because we use a completely consistent and an accurate numerical scheme,
which allows us to consider  the numerical errors as negligible.
Moreover, we will deal with the effect of using not-consistent Eulerian
models. 
Fianlly, a comparison between viscous and high-Reynolds results is carried out.
\section{Viscous Model}\label{sec:mod}

In this section, we propose a new Langevin model which account for 
the viscous terms present in the exact PDF equation.
Indeed, starting from the exact equations for a fluid,
Navier-Stokes equations
\begin{subequations}
\label{fluid: exact field Eqs.}
\begin{align}
&\frac{\partial U_{j}}{\partial x_j}=0, \\
&\frac{\partial U_{i}}{\partial t}+U_{j}\frac{\partial U_{i}}{\partial x_j
}
= -\frac{1}{\rho_f}\frac{\partial P}{\partial x_i} +
   \nu \frac{\partial^2 U_{i}}{\partial x_j^2} ,
\end{align}
\end{subequations}
we can derive an exact (yet not closed) equation for the one-particle 
PDF equation for the state-vector ${\bf{Y}}$
\begin{equation}\label{PDF_exact}
\begin{split}
\frac{\partial f}{\partial t} +
V_i\frac{\partial f}{\partial x_i} = &
-\frac{\partial}{\partial V_i}
[\, \langle -\frac{1}{\rho}\frac{\partial P}{\partial x_i}|
{\mathbf{Y}}={\mathbf{y}} \rangle\, f \,] \\
& -\frac{\partial}{\partial V_i}
[\, \langle \nu \Delta U_i |
{\mathbf{Y}}={\mathbf{y}} \rangle\, f \,] ,
\end{split}
\end{equation}
this equation may be considered as the starting point for stochastic modeling.
In high-Reynolds number flows, all viscous terms, 
but the turbulent dissipation, are generally neglected;
however, in low-Reynolds number flows or in wall-bounded flows,
they  can become necessary.
In order to include those cases also in the PDF approach, 
two different Langevin viscous models have been already 
proposed and tested \cite{Pop_97,Wac_04}.
Here we propose a third possible model for the state-vector $({\bf x},{\bf U})$:
\begin{eqnarray}
dx_i &=& U_i \,dt\label{modelp}\\
dU_i &=& -\frac{1}{\rho}\frac{\partial \langle P \rangle}{\partial x_i}\, dt +
\nu \frac{\partial^2 \langle U_i \rangle}{\partial x^2_k}\, dt \nonumber \\
& &+A_{il} (U_l-\lra{U_l})\, dt + G_{il} (U_l-\lra{U_l})\, dt \nonumber \\
&& + \sqrt{C_0\langle \epsilon \rangle}dW_i, \label{modelv}
\end{eqnarray}
where the matrix $A_{ij}$ is defined implicetely through the equation:
\begin{equation}
(AR)_{ij} = \frac{\nu}{2} \left(\frac{\partial^2 R_{ij}}{\partial x^2_k}\right)\,, \text{with}~ R_{ij} = \lra{u^{}_iu^{}_j};
\label{A-def}
\end{equation}
and the matrix ${G_{ij}}$ is the general matrix that models the pressure-fluctuation term.
In this work, we use the simplified Langevin model (SLM), that is
\begin{equation}
G_{ij} = - \left( \frac{1}{2} + \frac{3}{4}C_0 \right)
\frac{\langle \epsilon \rangle}{k}\;=\;-\frac{1}{T_L}~.
\end{equation}
At last, ${\bf{W}}$ is a Wiener process \cite{Ott_96}.

This model has been developed in the same framework of the two others cited above,
thus, for the general discussion of physical properties of this class of models
we refer to those papers. In the appendix \ref{app:reystress}, we show that the model proposed 
(\ref{modelp})-(\ref{modelv}) gives the correct equations for the moments of order 1
(mean-velocity),
 and 2 (Reynolds-stress) and, therefore, 
that the model is consistent with the known physics of fluid
turbulence.

The main specificity of this new model is that the equation (\ref{modelp}) for the position 
of the stochastic particles is maintained equal to the exact equation 
of fluid particle-position (in a Lagrangian sense).
On the contrary, both other models used the artifice of adding a white noise in the modeled
equation of particle position to represent the effect of diffusion.
Whilst this feature is only of formal relevance in fluid turbulence,
it can be useful in the case of two-phase flows \cite{Chi_08jas}, where
only this new model can be straightforwardly used. 
  Since the equation for particle position is constrained to 
remain not-modified, an ad-hoc term has been added in the modeled equation for particle velocity,
such that the diffusion in space for mean-velocity and
the term constituted by the matrix $A_{ij}$ is necessary to assure the correct equations for 
the Reynolds-stress.
In general flows, the matrix $A_{ij}$ can be computed from the formula (\ref{A-def}),
that can be rewritten in the form
\begin{equation}
{\bf A}\cdot {\bf R} = {\bf \epsilon}~; \text{hence}~ {\bf A} = {\bf \epsilon}\cdot {\bf R}^{-1}~.
\end{equation}

In present work, we consider the simple case of a channel flow, 
and further simplifications
can be made considering the symmetry properties of Reynolds-stress in x and z directions.
Furthermore, we make the hypothesis that the Reynolds-stress are nearly constant in the logarithmic zone,
which is well verified experimentally. This is, in fact, an equilibrium hypothesis and, 
on his basis, we can take just the leading terms in the viscous zone and
solve the equation analytically.
Then, we have
\begin{equation}
\epsilon = 
\begin{pmatrix}  \frac{\nu}{2} \frac{\partial^2 \lra{u^{2}}}{\partial x^2_i}&  0 & 0 \\
0 & 0 & 0\\
0 & 0 & \frac{\nu}{2} \frac{\partial^2 \lra{w^{2}}}{\partial x^2_i} \end{pmatrix}
\end{equation}
the resulting matrix $A_{ij}$ has the form
\begin{equation}
{\bf A} = \begin{pmatrix}  A_{11}& A_{12}  & 0 \\
0 & 0 & 0\\
0 & 0 & A_{33} \end{pmatrix}
\end{equation}
with
\begin{eqnarray}
A_{11} &=&  \frac{\nu}{2} \frac{\partial^2 \lra{u^{2}}}{\partial x^2_i} \frac{\lra{v^{2}}}
{\lra{u^{2}}\lra{v^{2}} - \lra{uv}^2} \\
A_{12} &=&  -\frac{\nu}{2} \frac{\partial^2 \lra{u^{2}}}{\partial x^2_i} \frac{\lra{uv}}
{\lra{u^{2}}\lra{v^{2}} - \lra{uv}^2} \\
A_{33} &=&  \frac{\nu}{2} \frac{\partial^2 \lra{w^{2}}}{\partial x^2_i} \frac{1}
{\lra{w^2}}~.
\end{eqnarray}

\section{Numerical Scheme}
Given the physical model, two main issues remain:
\begin{enumerate}
\item[(i)] Numerical integration scheme
\item[(ii)] Boundary conditions for particles
\end{enumerate}

\subsection{Stochastic differential system} \label{sec:sys}
The set of SDEs to be integrated can be written in a compact form
\begin{equation}
\left\{
\begin{split} \label{eq:sysEDS}
dx_{i} & =U_{i}\, dt, \\
dU_{i} & =-\frac{1}{T_i}U_{i}\,dt+ A_{12} U_2 \delta_{1,i} dt +C_i\, dt
         + \sum_j \sigma_{ij}\, dW_j(t),
\end{split} \right .
\end{equation}
where the diffusion matrix is diagonal 
and is defined
\begin{equation} \label{eq:sig}
\sigma _{ij}(t,{\bf x},\lra{{\bf U}})
= \lra{\epsilon}C_0\delta_{ij},
\end{equation}
and vector ${\bf C}$ is the sum of several terms
\begin{eqnarray} \label{eq:Ci}
C_1 &=&
\frac{\lra{U_{1}}}{T_1} - A_{12} \lra{U_2} -
\frac{1}{\rho}\frac{\partial \lra{P}}{\partial x}+ 
\nu\nabla^2\lra{U_{1}} \\
C_2 &=&
\frac{\lra{U_{2}}}{T_L}  -
\frac{1}{\rho}\frac{\partial \lra{P}}{\partial y}+ 
\nu\nabla^2\lra{U_{2}} \\
C_3 &=&
\frac{\lra{U_{3}}}{T_3}  -
\frac{1}{\rho}\frac{\partial \lra{P}}{\partial z}+ 
\nu\nabla^2\lra{U_{3}}~,
\end{eqnarray}
and the time-scales have become
\begin{equation} \label{def-time}
\frac{1}{T_i} = \frac{1}{T_L} - A_{ii}~.
\end{equation}

We recall that this model has a physical meaning only in
the case where $ \eta \ll dt \ll T_i$, with  $\eta$ the Kolmogorov time-scale.
When this condition
is not satisfied, it is possible to show that, in the
\textit{continuous sense} (time and all coefficients are continuous
functions which can go to zero), the system converges towards several
limit systems \cite{Min_00}:
\underline{case 1}: when $T_i \rightarrow 0$
with no condition on $\sigma_{ij}
T_i$, the velocity is no longer random and the system
becomes deterministic. The flow is laminar and it can be proven that
\begin{equation} \label{eq:limc1}
\text{system (\ref{eq:sysEDS}}) \;
\xrightarrow[T_i \to 0]{}
\left\{\begin{split}
& dx_{i} = U_{i}\, dt \\
& U_{i} = \lra{U_{i}}.
\end{split}\right.
\end{equation}
That is the expected behavior in the viscous sub-layer.

\underline{case 2}: When $T_i \rightarrow 0$ and at the same
time $\sigma_{ij}T_i \to cst$ and $A_{1,2}=cst$,
the fluid velocity becomes a fast variable. It can
be shown that
\begin{equation} \label{eq:limc2}
\text{system (\ref{eq:sysEDS})} \;
\xrightarrow[T_i \to 0]{(\sigma_{ij} T_i \to cst)}
dx_{i} = \lra{U_{i}}\,dt +
\sum_j (\sigma_{ij}T_i)\,dW_j(t). 
\end{equation}
We retrieve a pure diffusive behavior, that is the equations of the 
{\it Brownian motion}. 
Actually, this limit is formal in the continuous sense, 
being impossible to fit both conditions in a turbulent bounded flow 
($\sigma \sim \sqrt{\epsilon}$ which is always bounded).
Nevertheless, in numerical computations  this limit can be recovered
\cite{Min_03} and, furthermore, it has a theoretical relevance,
since those conditions are  required for the  elimination
of a variable, treated as fast variable.
The same reasoning has been applied to the fluid acceleration 
to obtain the present model on velocity \cite{Min_01}.
Thus, it is important that the numerical scheme is consistent even 
with this formal limit.

In our model, the diffusion matrix is diagonal, Eq. (\ref{eq:sig}).
Although following manipulations are strictly valid for the general case,
for the sake of simplicity, from now the diagonal expression
will be retained.

\subsection{Analytical solutions} \label{sec:sol}
The construction of the numerical schemes is now slightly
anticipated. Since the numerical methods are derived by freezing
the coefficients on the integration intervals, the solutions to system 
(\ref{eq:sysEDS}), \textit{with constant coefficients}, are now given.

The method of the constant variation is used. For the fluid velocity with $i\ne 1$,
 if one seeks a solution of the form $U_{i}(t)=H(t)\exp(-t/T_i)$, where $H(t)$
is a stochastic process, It\^o's calculus gives
\begin{equation} 
dH(t)=\exp(t/T_i)[C_i(\mb{x})\,dt +
       \sigma_{i}(\mb{x})\,dW_j(t)],
\end{equation}
that is, by integration on a time interval $[t_0,t]$ ($\Delta t=t-t_0$),
\begin{equation} 
\begin{split}
U_{i}(t) = U_{i}(t_0)\exp(-\Delta t/T_i)+C_i &
             (\mb{x})\,T_i\,[1-\exp(-\Delta t/T_i)] \\
+ & \sigma_{i}(\mb{x})\exp(-t/T_i)
    \int_{t_0}^{t}\exp(s/T_i)\,dW_i(s). 
\end{split}
\end{equation}
We can define
\begin{equation} 
\gamma_{i}(t) = \sigma_{i}(\mb{x})\exp(-t/T_i)
    \int_{t_0}^{t}\exp(s/T_i)\,dW_i(s)~; \text{for}~ i \ne 1
\end{equation}

For $i=1$ we substitute this expression to $U_2$ and we proceed in the same way to
obtain
\begin{equation} 
\begin{split}
& U_{1}(t) =  ~U_{1}(t_0)\exp(-\Delta t/T_1)+C_1(\mb{x})\,T_1\,[1-\exp(-\Delta t/T_1)] \\
&+  A_{12}\exp(-t/T_1)\int_{t_0}^{t}\exp(s/T_1)
\left[U_{2}(t_0)\exp(-\Delta s/T_2)+C_2 
             (\mb{x})\,T_2\,(1-\exp(-\Delta s/T_2))
+\gamma_2(s)\right] ds \\
&+\sigma_{1}(\mb{x})\exp(-t/T_1)
    \int_{t_0}^{t}\exp(s/T_1)\,dW_1(s). 
\end{split}
\end{equation}
Performing integration by parts we, finally, obtain
\begin{eqnarray}
U_{1}(t)&=& U_{1}(t_0)\exp(-\Delta t/T_1+
[C_1 (\mb{x})\,T_1+C_2 (\mb{x})\,T_2A_{12}T_2]\left(1-\exp(-\Delta t/T_1)\right)
\nonumber \\
&+&A_{12}\theta[ U_{2}(t_0) - C_2 (\mb{x})\,T_2]
\left[\exp(-\Delta t/T_2) -\exp(-\Delta t/T_1)\right] + \gamma_1(t) 
\end{eqnarray}
having defined
\begin{eqnarray}
\gamma_1(t) &=& A_{12} \exp(-t/T_1)\int_{t_0}^{t}\exp(s/T_1)\gamma_2(s)ds +
\sigma_1\exp(-t/T_1)\int_{t_0}^{t}\exp(s/T_1) dW_1 \\
\theta&=& \frac{T_1T_2}{T_2-T_1}~.
\end{eqnarray}
The first integral in the definition of $\gamma_1$ is a double stochastic integral 
which can be simplified through the integration by parts,
 as explained in appendix \ref{app:intpart}.
The new formula is
\begin{eqnarray} 
\gamma_1(t)&=& A_{12} \sigma_2 \theta
 \left[\exp(-t/T_2)\int_{t_0}^{t}\exp(s/T_2) dW_2 -
\exp(-t/T_1)\int_{t_0}^{t}\exp(s/T_1) dW_2\right] \nonumber \\
&+&\sigma_1\exp(-t/T_1)\int_{t_0}^{t}\exp(s/T_1) dW_1 
\end{eqnarray}

The analytical solutions for the positions are worked out is a similar manner
starting from Eq. (\ref{modelp})
\begin{equation} 
dx_i = U_i dt \Rightarrow x_i(t) = x_i(t_0) + \int_{t_0}^{t}U_i(s)ds
\end{equation}
for $i\ne1$
\begin{eqnarray}
x_{i}(t) &=& x_{i}(t_0)+U_{i}(t_0)T_i [1-\exp(-\Delta t/T_i)]
+[C_i(\mb{x})T_i]\{\Delta t-T_i[1-\exp(-\Delta t/T_i)]\} +\Gamma_1(t) \\
\Gamma_i(t) &=& \sigma_i\int_{t_0}^{t}\exp(-t^{\prime}/T_i)
\int_{t_0}^{t^{\prime}}\exp(s/T_i)dW_i(s)
\end{eqnarray}
As usual, the double integral can be transformed
\begin{equation}
\Gamma_i(t) = \sigma_iT_i \left[\int_{t_0}^{t}dW_i(s)-\exp(-t/T_i)
\int_{t_0}^{t}\exp(s/T_i)dW_i(s)\right]~;~\text{for} ~i=\{2,3\}~.
\end{equation}

Finally, the integration of the first component in position gives
\begin{eqnarray}
x_{1}(t) &=& x_{1}(t_0)+U_{1}(t_0)T_1 [1-\exp(-\Delta t/T_1)]
+T_1[C_1(\mb{x})+A_{12}C_2(\mb{x})T_2]\times \nonumber \\
&\times &\{\Delta t-T_1[1-\exp(-\Delta t/T_1)]\} 
+ A_{12} \theta [U_2(t_0)-C_2(\mb{x})T_2]\times \nonumber \\
& \times & [T_2(1-\exp(-\Delta t/T_2))
-T_1(1-\exp(-\Delta t/T_1))]+ \Gamma_1(t) \\
\Gamma_1(t) &=& A_{12} \sigma_2 \theta \left\{(T_2-T_1)\int_{t_0}^{t}dW_2(s) 
-T_2\exp(-t/T_2)\int_{t_0}^{t}\exp(s/T_2)dW_2(s) \right.\nonumber \\ 
&+& \left. T_1\exp(-t/T_1)\int_{t_0}^{t}\exp(s/T_1)dW_2(s)\right\}\nonumber \\ 
&+& \sigma_1\left[\int_{t_0}^{t}dW_1
-T_1\exp(-t/T_1)\int_{t_0}^{t}\exp(s/T_1)dW_1(s)\right]~;
\end{eqnarray}
also for the position the stochastic integral $\Gamma_1$ has been reduced 
in his basic components through integration by parts.
To conclude this section, we resume all the analytical solutions in table
\ref{tab:exa}.

The stochastic integrals, Eqs (\ref{Gamma1}) to
(\ref{gammai}) in Table \ref{tab:exa}, only for the constant coefficient
case, are Gaussian processes
since they are stochastic integrals of deterministic functions
\cite{Kle_99}.
These integrals are quite involved, but, actually, they are expressed 
in function of some simple stochastic integrals which can be used as base.
The base is formed by the following 7 integrals
\begin{eqnarray}
I_{1,i} &=& \exp(-t/T_i)\int_{t_0}^{t}\exp(s/T_i)dW_i(s)\quad i=\{1,2,3\} \label{I1i}\\
I_{2,2} &=& \exp(-t/T_1)\int_{t_0}^{t}\exp(s/T_1)dW_2(s) \\
I_{3,i} &=& \int_{t_0}^{t}dW_i(s) \quad i=\{1,2,3\}~.\label{I3i}
\end{eqnarray}
The original stochastic integrals eqs (\ref{Gamma1}) to
(\ref{gammai}) are, then, decomposed and reformulated as
\begin{eqnarray}
\text{i$\ne$1} \quad \quad  \gamma_i &=& \sigma_i I_{1,1} \\
\text{i=1} \quad \quad \gamma_1 &=& A_{12} \sigma_2 \theta[I_{1,2} -I_{2,2}] 
+ \sigma_1 I_{1,1} \\
\nonumber \\
\text{i$\ne$1} \quad \quad \Gamma_i &=& \sigma_i T_i [I_{3,1} -I_{1,i}]\\
\text{i=1} \quad \quad \Gamma_1 &=& A_{12} \sigma_2 \theta[(T_2-T_1)I_{3,2} -
T_2I_{1,2}+T_1I_{2,2}] + \sigma_1 T_1[I_{3,1} -I_{1,1}] ~.
\end{eqnarray}
In a matrix form the two set of integrals are related by the formula
\begin{equation}\label{y=mx}
\begin{pmatrix} \gamma_1 \\ 
	 \vdots \\
	 \Gamma_3 \end{pmatrix} = {\bf M} \cdot  
\begin{pmatrix} I_{1,1} \\ 
	 \vdots \\
	 I_{3,3} \end{pmatrix}
\end{equation}
with 
\begin{equation} \label{matriceM}
{\bf M} = \begin{bmatrix}
\sigma_1 & A_{12}\sigma_2 \theta & 0 & -A_{12}\sigma_2 \theta&  & \dots  & 0 \\
0 & \sigma_2 & 0 & \dots &  & \dots & 0\\
0 & 0 & \sigma_3 &  & \dots &  & 0 \\
-\sigma_1T_1 & - T_2A_{12}\sigma_2 \theta & 0 & T_1 A_{12}\sigma_2 \theta &
\sigma_1T_1 & (T_2-T_1) A_{12}\sigma_2 \theta & 0 \\
0 & -\sigma_2 T_2 &0 & 0 & 0& \sigma_2 T_2 &0 \\
0 & 0 & -\sigma_3 T_3 &0 & 0 & 0 & \sigma_3 T_3 \end{bmatrix}
\end{equation}

In order to evaluate and to simulate numerically
stochastic integrals, 
it is necessary to compute the moments of integrals
and in particular the covariance matrix $C_{ij}$.
Here, we do not proceed to the complicated calculation of the covariance matrix 
of the integrals $\Gamma$ and $\gamma$, but we compute the simpler 
covariance matrix of the base integrals (\ref{I1i})-(\ref{I3i}).
Once computed this covariance matrix and with the help of 
the relations given by Eqs. (\ref{y=mx})-(\ref{matriceM}), it always possible 
to retrieve the covariance of $\Gamma$ and $\gamma$ and, then, 
to simulate them numerically.
The covariance matrix of the base integrals (\ref{I1i})-(\ref{I3i})
is obtained using the techniques explained 
in appendix \ref{app:covmat} and the resulting values are shown 
in table \ref{tab:matcov_exa}.

Again, we slightly anticipate the numerics and it is 
explained how the stochastic integrals can be calculated
(simulated). It has just been shown that these integrals are Gaussian
processes whose means and variances are know (zero mean and covariance
matrix given by Table \ref{tab:matcov_exa}). The vector composed by
the seven stochastic integrals, eqs (\ref{I1i}) to
(\ref{I3i}), is a vector of Gaussian centered random
variables which can be computed by resorting to the simulation of a
vector composed of independent normal Gaussian variables (zero mean
and variance equal to one). This technique, which requires the
Choleski decomposition of the covariance matrix, is displayed in
Appendix \ref{app:choleski}.
It is worth recalling that once computed the stochastic integrals forming our base,
the integrals (\ref{Gamma1})-(\ref{gammai}) can be obtained 
through the matrix relation (\ref{y=mx}).

At last, let us check that the expressions of Table \ref{tab:exa} and
\ref{tab:matcov_exa} are consistent with the limit cases given,
\textit{in the continuous sense}.

\underline{case 1}: when $T_i \rightarrow 0$
with no condition on $\sigma_{i}$, the
flow becomes laminar, which means that the system becomes
deterministic, see Eq. (\ref{eq:limc1}). Once again, the results given 
by Table \ref{tab:exa} and \ref{tab:matcov_exa} are in agreement with
Eq. (\ref{eq:limc1}). When $T_i\rightarrow 0$, eqs (\ref{eq:x1_exa})
to (\ref{eq:Ui_exa}) become
\begin{equation} 
\begin{split}
& U_{i}(t)=\lra{U_{i}(t)}, \\
& x_{i}(t) = x_{i}(t_0) + \lra{U_{i}(t)} \Delta t ,
\end{split}
\end{equation}
which is the analytical solution to system (\ref{eq:limc2}) when the
coefficients are constant.

\underline{case 2}: When $T_i \rightarrow 0$ and at the same
time $\sigma_{i}T_i \to cst$,
we have 
\begin{equation} 
dx_{i} = \lra{U_{i}}\,dt +
\sigma_{i}T_i \sqrt{\Delta t} {\mc{G}}_i~
\end{equation}
where ${\mc G}_{i}$ is ${\mc N}(0,1)$ vector (a vector composed of
independent normal Gaussian random variables).

A major conclusion can be derived, 
in order to have a consistent and stable numerical scheme for the integration,
even within the asymptotic limits,
it is necessary to retain the exponential terms in the solution
and all the stochastic integrals, even the {\it indirect} ones, 
must be explicitly calculated.

\begin{table}[!h]
\begin{center}
\caption{\small Analytical solutions to system (\ref{eq:sysEDS}) for
time-independent coefficients.}
\hrule
\begin{align}
x_{1}(t)& = x_{1}(t_0)+U_{1}(t_0)T_1 [1-\exp(-\Delta t/T_1)]
+T_1[C_1(\mb{x})+A_{12}C_2(\mb{x})T_2]\times \notag \\
&\times \{\Delta t-T_1[1-\exp(-\Delta t/T_1)]\} 
+ A_{12} \theta [U_2(t_0)-C_2(\mb{x})T_2]\times \notag \\
&\times [T_2(1-\exp(-\Delta t/T_2))-T_1(1-\exp(-\Delta t/T_1))] + \Gamma_1(t) \label{eq:x1_exa}\\
x_{i}(t) &= x_{i}(t_0)+U_{i}(t_0)T_i [1-\exp(-\Delta t/T_i)]
+[C_i(\mb{x})T_i]\{\Delta t-T_i[1-\exp(-\Delta t/T_i)]\} \notag \\
& +\Gamma_1(t) \label{eq:xi_exa}\\ 
U_{1}(t)&= U_{1}(t_0)\exp(-\Delta t/T_1+
[C_1 (\mb{x})\,T_1+C_2 (\mb{x})\,T_2A_{12}T_2]\left(1-\exp(-\Delta t/T_1)\right)
\notag \\
&+A_{12}\theta[ U_{2}(t_0) - C_2 (\mb{x})\,T_2]
\left(\exp(-\Delta t/T_2 -\exp(-\Delta t/T_1\right) + \gamma_1(t) \label{eq:U1_exa}\\
U_{i}(t) &= U_{i}(t_0)\exp(-\Delta t/T_i)+C_i(\mb{x})\,T_i\,[1-\exp(-\Delta t/T_i)] \notag \\
&  + \sigma_{i}(\mb{x})\exp(-t/T_i) \int_{t_0}^{t}\exp(s/T_i)\,dW_i(s) \label{eq:Ui_exa}\\ 
\notag \\
& \text{\underline{The stochastic integrals $\gamma _i(t),\;\Gamma
_i(t),$ are given by:}}\notag \\
\Gamma_1(t) &= A_{12} \sigma_2 \theta \left[(T_2-T_1)\int_{t_0}^{t}dW_2(s)
-T_2\exp(-t/T_2)\int_{t_0}^{t}\exp(s/T_2)dW_2(s) \right. \notag \\
\quad &+\left. T_1\exp(-t/T_1)\int_{t_0}^{t}\exp(s/T_1)dW_2(s)\right] \notag \\
\quad &+\sigma_1\left[\int_{t_0}^{t}dW_1
-T_1\exp(-t/T_1)\int_{t_0}^{t}\exp(s/T_1)dW_1(s)\right]  \label{Gamma1}\\
\Gamma_i(t) &= \sigma_iT_i \left[\int_{t_0}^{t}dW_i(s)-\exp(-t/T_i)
\int_{t_0}^{t}\exp(s/T_i)dW_i(s)\right]~;~\text{for} ~i=\{2,3\} \label{Gammai} \\
\notag \\
\gamma_1(t) &= A_{12} \exp(-t/T_1)\int_{t_0}^{t}\exp(s/T_1)\gamma_2(s)ds +
\sigma_1\exp(-t/T_1)\int_{t_0}^{t}\exp(s/T_1) dW_1 \label{gamma1}\\
\gamma_{i}(t) &= \sigma_{i}(\mb{x})\exp(-t/T_i)
    \int_{t_0}^{t}\exp(s/T_i)\,dW_i(s) \label{gammai}
\end{align}
\hrule
\label{tab:exa}
\end{center}
\end{table}

\clearpage
\begin{table}[h]
\begin{center}
\caption{\small Derivation of the covariance matrix for constant coefficients.}
\hrule
\begin{align}
& \lra{I_{1,i}^2(t)}  = \frac{T_i}{2} \left[1-\exp(-2\Delta t/T_i)\right] 
\label{eq:I1i2} \\ \notag \\
& \lra{I_{2,2}^2(t)} =
\frac{T_1}{2} \left[1-\exp(-2\Delta t/T_1)\right]
\label{eq:I222} \\ \notag \\
& \lra{I_{3,i}^2(t)} = \Delta t
\label{eq:I3i2} \\ \notag \\
& \lra{I_{1,i}I_{3,i}} = T_i\left[1-\exp(-\Delta t/T_i)\right] 
\label{eq:I132} \\ \notag \\
& \lra{I_{1,2}I_{2,2}} = \frac{T_2T_1}{T_2+T_1}
\left[1-\exp(-\Delta t/T_1)\exp(-\Delta t/T_2)\right] 
\label{eq:I122} \\ \notag \\
& \lra{I_{2,2}I_{3,2}} = T_1\left[1-\exp(-\Delta t/T_1)\right] 
\label{eq:I232} \\ \notag 
\end{align}
\hrule
\label{tab:matcov_exa}
\end{center}
\end{table}

\begin{table}[h]
\caption{\small Weak first order scheme (Euler scheme): sch1}
\hrule
\begin{align}
& \text{\underline{Numerical integration of the system}}:\notag \\
& \text{i$\ne$1} \quad x_{i}^{n+1} = x_{i}^n + A\,U_{i}^n + B\,T_i^n C_i^n
 + \Gamma_i^n, \notag \\
& \text{i=1} \quad x_{1}^{n+1} = x_{1}^n + A\,U_{1}^n + B\,T_1^n 
[C_1^n+A_{12}^nC_2^nT-2^n] + C\, A_{12}^n[U_2^n - C_2^n T _2^n]+ \Gamma_1^n, \notag \\
  \notag \\
& \text{i$\ne$1} \quad U_{i}^{n+1} = U_{i}^n\, \exp(-\Delta t/T_i^n)
              + [T_i^n C_i^n] D [1-\exp(-\Delta t/T_i^n)]
              + \gamma _i^n, \notag \\
& \text{i=1} \quad U_{i}^{n+1} = U_{i}^n\, \exp(-\Delta t/T_1^n)
              + T_1^n [C_1^n + A_{12}^nC_2^n T _2^n] D 
              + [U_2^n - C_2^n T _2^n]E
              + \gamma _i^n, \notag \\
 \notag \\ \notag \\
& \text{\underline{The coefficients $A,\;B,\;C,\;D$ and $E$ are
  given by}}:\notag \\ 
& \quad A = T_i^n(1-\exp(-\Delta t/T_i^n) \notag \\
& \quad B = \Delta t - A \notag \\
& \quad C = \theta^n [T_2^n(1-\exp(-\Delta t/T_2^n)) - T_1^n(1-\exp(-\Delta t/T_1^n))]
\quad \text{with} \quad \theta^n = \frac{T_2^nT_1^n}{T_2^n-T_1^n} \notag \\
& \quad D = [1-\exp(-\Delta t/T_i^n)] \notag \\
& \quad E = A_{12}^n \theta^n \left(\exp(-\Delta t/T_2^n)-\exp(-\Delta t/T_1^n)\right) \notag \\
\end{align}
\hrule
\label{tab:sch1}
\end{table}

\subsection{Constraints of the numerical schemes} \label{sec:cont}
In the particle-mesh method adopted here, the PDEs for the fluid are
first solved and then the dynamics of the stochastic particles is
computed. Thus, the scheme has to be \textit{explicit}. Furthermore,
the time step, which should be the same for the integration of the
PDEs and the SDEs,  is imposed by stability conditions required by the
Eulerian integration operator. This implies that, since there is no possibility to
control the time step when integrating the SDEs, the numerical scheme
has to be \textit{unconditionally stable}. At last, since particle localization in a
mesh is CPU demanding, the numerical scheme should minimize these
operations. The first constraint is
\begin{enumerate}
\item[(i)] \textit{The numerical scheme must be explicit, stable, of
order $2$ in time and the number of calls to particle localization has
to be minimum}.
\end{enumerate}

A practical strategy to fulfill the stability condition is to base the scheme
on the analytical solutions presented in Section
\ref{sec:sol}. Indeed, the time step appears in decaying exponentials
of the  type $\exp(-\Delta t)$, which brings unconditional
stability. Therefore, the second constraint is
\begin{enumerate}
\item[(ii)] \textit{ The numerical scheme must be consistent with the
analytical solutions of the system when the coefficients are constant}.
\end{enumerate}
\begin{enumerate}
\item[(iii)] \textit{The numerical scheme must be consistent with all
limit systems}.
\end{enumerate}

For a general and comprehensive discussion about 
the delicate point of asymptotic limit,
we refer to the paper \cite{Pei_06}, 
where this issue is analyzed in the more general case of
two-phase flows.
The construction of the scheme on the analytical solutions should also 
ensure a sound physical treatment of the multi-scale character of the
problem. Indeed, it has been demonstrated that from the analytical
solutions given in Section \ref{sec:sol}, the limit cases can be
retrieved.

\subsection{Weak first-order scheme} \label{sec:sch1}

The derivation of the weak first order scheme is rather straightforward 
since the analytical solutions of system (\ref{eq:sysEDS})
with constant coefficients have been evaluated. Indeed, the Euler
scheme (which is a weak scheme of order $1$ \cite{Klo_92}) is simply
obtained by freezing the coefficients at the beginning of the time
intervals $\Delta t = [t_n,t_{n+1}]$. Let $Z_i^n$ and $Z_i^{n+1}$  be
the approximated values of $Z_i$ at time $t_n$ and $t_{n+1}$. The
Euler scheme is then simply written by using the results of Tables
\ref{tab:exa} and \ref{tab:matcov_exa} as shown in Table
\ref{tab:sch1}.
This scheme fulfills all criteria listed in Section \ref{sec:cont}, except,
of course, the order of convergence in time.

As far as the consistency with all limit systems is concerned, some
precisions must be given. Here, it has to be understood that the limit
systems are presented in the \textit{discrete sense}. The observation
time scale $dt$ has now become the time step $\Delta t$. The physical
time scale $T_i$ does not go to zero, as in the continuous
sense, but its value, depending on the history of the particles, can
be smaller or greater than $\Delta t$.

In the limit case 1 (\ref{eq:limc1}) we have, with $1 \ll \Delta t/T_i$,

\begin{eqnarray} 
\Gamma_i(t) &&\xrightarrow[T_i \to 0]{}
\sigma_i\,T_i \sqrt{\Delta t} {\mc G}_{x,i} \approx 0 \\
\gamma_i(t) &&\xrightarrow[T_i \to 0]{} 0
\end{eqnarray}
and, thus, for the fluid variables 
\begin{eqnarray} 
U_i^{n+1}&=&\lra{U_i^n} \\
x^{n+1}_i&=&x_i^n+\lra{U_i^n} \Delta t~.
\end{eqnarray}
In this case these relations are identical to those analytic.
In the second limit case (\ref{eq:limc2}) it is different;
for the present scheme, Eq. (\ref{eq:Ui_exa}),
with $1 \ll \Delta t/T_i$ and the constraints on $\sigma$, gives
\begin{equation} 
U_{i}^{n+1} = \lra{U_{i}^{n}} +
\sqrt{\frac{\sigma_i^{n2}\,T_i^n}{2}}\;\;{\mc G}_{1,i},
\end{equation}
From the continuous point of view, the fluid velocity  $U_i$ will
behave as a fast variable and a near white-noise term.
On the contrary, in the numerical simulation, the velocity 
$U_i$ does not, of course, behave strictly-speaking as white-noise
function (its variance is not infinite!).
This result is physically sound. Indeed, when a fluctuating physical
process (random variable) is observed at time steps which are greater
that its memory, the expected behavior is Gaussian.
For particle position the scheme gives
\begin{equation} 
x^{n+1}_{i} = x^n_{i} + \lra{U^n_{i}} \Delta t +
+ \sqrt{\sigma_i^{n2}\,T_i^{n2}\,\Delta t} \;\;{\mc G}_{x,i},
\end{equation}
as expected.
\subsection{Weak second order scheme}

\subsubsection{Property of the system} \label{sec:prop}
The diffusion matrix of system (\ref{eq:sysEDS}) has a singular
property, that has a crucial importance here \cite{Min_03}.
 From Eq. (\ref{eq:sig}), it can be noticed that
$\sigma_{ij}$ depends only on time, position and the mean value of the 
relative velocity. Therefore, the only variable of the state vector
whose $\sigma_{ij}$ is a function is ${\mb x}_p$. Therefore, the
diffusion matrix has the following singular property

\begin{equation}\label{sig-prop}
\sum_k \sum_j \sigma_{kj}\,\frac{\partial \sigma_{ij}}{\partial
x_k}=0, \quad \forall \; (i).
\end{equation}

\subsubsection{General method} \label{sec:idee}
Let us consider, for a moment, the following stochastic differential
equation
\begin{equation} \notag
dX_i(t) = A_i(\mb{X}(t))\,dt + \sum_j B_{ij}(\mb{X}(t))\,dW_j(t)
\end{equation}
where ${\mb A}$ is the drift vector and ${\mb B}$ is the diffusion
matrix. If ${\mb B}$ verifies the property (\ref{sig-prop}), it
can be shown that, for example by stochastic Taylor expansions
\cite{Klo_92}, a prediction-correction scheme of the type
\begin{equation} \label{eq:idee}
\begin{split}
& \td{X}_i^{n+1} = X_i^n + A_i^n\,dt + \sum_j B_{ij}^n\,\Delta W_j, \\
& X_i^{n+1} = X_i^n + \frac{1}{2}\left(A_i^n+\td{A}_i^{n+1}\right)\,dt
 + \sum_j \frac{1}{2}\left(B_{ij}^n+\tilde{B}_{ij}^{n+1}\right)\,\Delta W_j,
\end{split}
\end{equation}
is a weak second order scheme ($\td{A}_i^{n+1}=A_i(\td{\mb{X}}^{n+1})$
and $\td{B}_{ij}^{n+1}=B_{ij}(\td{\mb{X}}^{n+1})$). This result is
true, one again, only when the property (\ref{sig-prop}) is verified. If 
not, other terms are needed in order to obtain a weak second order
scheme, see for example Talay \cite{Tal_95}.
As a consequence, it this scheme is used for a set of stochastic equations
which do not verify the property (\ref{sig-prop}),
problems of consistency arise \cite{Min_03a}.

The idea is now to  built from Eq. (\ref{eq:idee})  a weak
second order scheme for system (\ref{eq:sysEDS}).
The first-order Euler scheme  is
going to be used as a prediction step. The remaining task consists in
finding a suitable correction step which enforces the three conditions
listed in Section \ref{sec:cont}.

\subsubsection{Implementation of the scheme}
The main idea for the implementation of the correction step is to
start from the analytical solutions to system (\ref{eq:sysEDS}) when
the coefficients are constant and apply the same idea as in
Eq. (\ref{eq:idee}) considering that the acceleration terms vary 
linearly with time.

Before the implementation of the numerical scheme, let us specify the
notation which is in use. As in Section \ref{sec:sch1}, $Z_i^n$
and $Z_i^{n+1}$  stand for the approximated values of $Z_i$ at time
$t_n$ and $t_{n+1}$, respectively. Therefore, $Z_i^{n+1}$ is the
corrected value whereas the predicted value is denoted
$\td{Z}_i^{n+1}$, \textit{i.e.} the value of $Z_i$ predicted by the
Euler scheme. The predicted velocities are 
$\td{U}_{i}^{n+1}$ and the values of the fields taken at
$(t_{n+1},\,\mb{x}^{n+1})$ are denoted, for example,
 $\lra{P^{n+1}}$. The predicted time scales
are referred to as $\td{T}_i^{n+1}$  and the
predicted diffusion matrix is designated by $\td{\sigma}_{i}^{n+1}$.

As far as the computation of the fields 'attached' to the particles
are concerned, it is worth emphasizing that none of them are
computed at $(t_{n+1},\,\mb{x}^{n+1})$, because the scheme would
become implicit. All fields which characterize the fluid, such as the
mean pressure field, are taken at $(t_{n+1},\,\mb{x}^{n+1})$, but
fields such as the expected value of the fluid velocity  are
computed from the predicted velocities. For example, one has
\begin{equation} \notag
C_i(t_{n+1},\,\mb{x}^{n+1}) = {\td{C}_i^{n+1}}=
  \frac{1}{\td{T}_i^{n+1}}\lra{U_{i}^{n+1}}
+ \Pi _i(\lra{\mb{U}^{n+1}},\lra{P^{n+1}} ).
\end{equation}

The analytical solution to system (\ref{eq:sysEDS}) when the
coefficients are constant is, for the fluid velocity seen, by applying 
the rules of It\^o's calculus
\begin{equation} \label{eq:Ui_sch2}
\text{i$\ne$1} \quad U_{i}(t) = U_{i}(t_0)\,\exp(-\Delta t/T_i)
+ \int _{t_0}^{t}C_i(s)\exp[(s-t)/T_i]\,ds + \gamma_i(t).
\end{equation}
As anticipated we suppose that the acceleration
$C_i(s,\mb{x}_p)$ varies linearly on the integration interval
$[t_0,t]$, that is
\begin{equation} \label{eq:Ci_int}
C_i(s,\mb{x}(s)) = C_i(t_0,\mb{x}(t_0)) + \frac{1}{\Delta t}
[C_i(t,\mb{x}(t))-C_i(t_0,\mb{x}(t_0))](s-t_0).
\end{equation} 
By direct integration, one can write
\begin{equation} \notag
U_{i}(t) = U_{i}(t_0)\,\exp(-\Delta t/T_i)
+ [T_i\,C_i(t_0,\mb{x}(t_0))]\,A(\Delta t,T_i)
+ [T_i\,C_i(t,\mb{x}(t))]\,B(\Delta t,T_i)+ \gamma_i(t),
\end{equation}
where the functions $A(\Delta t,X)$ and $B(\Delta t,X)$ are given by
\begin{equation} 
\begin{split}
& A(\Delta t,X) = -\exp(-\Delta t/X)+
 \left[\frac{1-\exp(-\Delta t/X)} {\Delta t/X}\right],\\
& B(\Delta t,X) = 1-\left[\frac{1-\exp(-\Delta t/X)}{\Delta
  t/X}\right]. 
\end{split}
\end{equation}
By direct application of the ideas presented in the last section,
 it is proposed to write
\begin{equation} 
\begin{split}
U_{i}^{n+1} = & 
\frac{1}{2}\,U_{i}^n\,\exp(-\Delta t/T_i^n)
+ \frac{1}{2}\,U_{i}^{n}\,\exp(-\Delta t/\td{T}_i^{n+1}) \\
+ & A(\Delta t,\,T_i^n) \,[T_i^n C_i^n]
+ B(\Delta t,\,\td{T}_i^{n+1}) \,[\td{T}_i^{n+1} \td{C}_i^{n+1}]
+ \td{\gamma}_i^{n+1},
\end{split}
\end{equation}
The meaning of the corrected stochastic integrals will be given later.
For the first components the exact solution is
\begin{equation} \label{eq:U1_sch2}
\begin{split}
\text{i$=$1} \quad U_{1}(t) = & U_{1}(t_0)\,\exp(-\Delta t/T_1)
+ \int _{t_0}^{t}C_1(s)\exp[(s-t)/T_1]\,ds  \\
+& \int _{t_0}^{t} A_{12}U_2(s)\exp[(s-t)/T_1]ds +
\int _{t_0}^{t} \exp[(s-t)/T_1]\sigma_1(s)dW_1(s)~.
\end{split}
\end{equation}
Here, we introduce the exact expression for the variable $U_2$
\begin{equation} \label{eq:U2_sch2}
U_{2}(t) = U_{2}(t_0)\,\exp(-\Delta t/T_2)
+ \int _{t_0}^{t}C_i(s)\exp[(s-t)/T_2]\,ds + \gamma_2(t).
\end{equation}
and, then, we let the functions $C_i$ and $\sigma$ varying linearly.
The subsequent integration, quite involved, permit us to propose 
the following expression
\begin{equation} 
\begin{split}
U_{1}^{n+1} = & 
\frac{1}{2}\,U_{1}^n\,\exp(-\Delta t/T_1^n)
+ \frac{1}{2}\,U_{1}^{n}\,\exp(-\Delta t/\td{T}_1^{n+1}) 
+ \frac{1}{2}U_{2}^nA_{12}^nE(T_1^n,T_2^n)\\
+& A(\Delta t,\,T_1^n) \,[T_1^n C_1^n] + 
B(\Delta t,\,\td{T}_1^{n+1}) \,[\td{T}_1^{n+1} \td{C}_1^{n+1}] \\
+& A_{12}^n[T_2^n C_2^n]A_c(T_1^n,T_2^n) 
+\td{A}_{12}^n[\td{T}_2^{n+1} \td{C}_2^{n+1}]B_c(T_1^n,T_2^n) 
+ \td{\gamma}_1^{n+1}~.
\end{split}
\end{equation}
where the functions  $A_c(X,Y)$ and $B_c(X,Y)$ are given by
\begin{equation} \notag
\begin{split}
& A_c(X,Y) =
- \exp(-\Delta t/X) + \frac{X+Y}{\Delta t}[1-\exp(-\Delta t/X)]
- \left( 1+\frac{Y}{\Delta t} \right)\,C_c(X,Y), \\
& B_c(X,Y) =  1 - \frac{X+Y}{\Delta t}[1-\exp(-\Delta t/X)]
+ \frac{Y}{\Delta t}\,C_c(X,Y).\\
&\text{with} \notag \\
& C_c(X,Y) = \frac{Y}{Y-X}\,[\exp(-\Delta t/Y)-\exp(-\Delta t/X)], \\
\end{split}
\end{equation}
\begin{table}[!h]
\caption{\small Weak second order scheme: sch2c}
\hrule
\begin{align}
& \text{\underline{Prediction step:}} \notag \\
&  \text{apply the Euler scheme Sch1,
see Table \ref{tab:sch1}}.\notag \\ \notag \\
& \text{\underline{Correction step:}} \notag \\
&U_{1}^{n+1}  =  
\frac{1}{2}\,U_{1}^n\,\exp(-\Delta t/T_1^n)
+ \frac{1}{2}\,U_{1}^{n}\,\exp(-\Delta t/\td{T}_1^{n+1}) 
+ \frac{1}{2}U_{2}^nA_{12}^nE(T_1^n,T_2^n) \notag \\
\quad&+ A(\Delta t,\,T_1^n) \,[T_1^n C_1^n] + 
B(\Delta t,\,\td{T}_1^{n+1}) \,[\td{T}_1^{n+1} \td{C}_1^{n+1}] \notag  \\
\quad&+ A_{12}^n[T_2^n C_2^n]A_c(T_1^n,T_2^n) 
+\td{A}_{12}^n[\td{T}_2^{n+1} \td{C}_2^{n+1}]B_c(T_1^n,T_2^n) 
+ \td{\gamma}_1^{n+1}~.\notag \\
\notag \\
&U_{i}^{n+1}  =  
\frac{1}{2}\,U_{i}^n\,\exp(-\Delta t/T_i^n)
+ \frac{1}{2}\,U_{i}^{n}\,\exp(-\Delta t/\td{T}_i^{n+1}) \notag \\
\quad&+  A(\Delta t,\,T_i^n) \,[T_i^n C_i^n]
+ B(\Delta t,\,\td{T}_i^{n+1}) \,[\td{T}_i^{n+1} \td{C}_i^{n+1}]
+ \td{\gamma}_i^{n+1}\notag \\
\notag \\
& \text{\underline{The coefficients $A,\;B,\;A_c,\;B_c$ et $C_c$ are
  defined as}}:\notag \\ 
&  A(\Delta t,X)  = -\exp(-\Delta t/X)
+ \left[\frac{1-\exp(-\Delta t/X)}{\Delta t/X}\right], \notag\\
&  B(\Delta t,X)  = 1-\left[\frac{1-\exp(-\Delta t/X)}{\Delta
  t/X}\right],\notag \\ 
&  A_c(X,Y)  =
- \exp(-\Delta t/X) + \frac{X+Y}{\Delta t}[1-\exp(-\Delta t/X)]
- \left( 1+\frac{Y}{\Delta t} \right)\,C_c(X,Y), \notag \\
&  B_c(X,Y) =  1 - \frac{X+Y}{\Delta t}[1-\exp(-\Delta t/X)]
+ \frac{Y}{\Delta t}\,C_c(X,Y), \notag\\
&  C_c(X,Y)  = \frac{Y}{Y-X}\,[\exp(-\Delta t/Y)-\exp(-\Delta
t/X)].\notag \\ \notag \\  
& \text{\underline{The stochastic integrals
        $\td{\gamma}_i^{n+1}\;$ \text{and} $\;\td{\Gamma}_i^{n+1}$
        are simulated as follows}}:\notag \\
&\gamma_i^{n+1} = [\sigma_i^n A_{\gamma}(2\Delta t,\td{T}_i^{n+1}) +
\sigma_i^{n+1} B_{\gamma}(2\Delta t,\td{T}_i^{n+1})] I_{1,i}(\td{T}_i^{n+1}) 
\quad \text{i=\{2,3\}}\notag\\
&\gamma_1^{n+1} = [\sigma_1^n A_{\gamma}(2\Delta t,\td{T}_1^{n+1}) +
\sigma_1^{n+1} B_{\gamma}(2\Delta t,\td{T}_1^{n+1})] I_{1,1}(\td{T}_i^{n+1}) 
\quad \text{i=1} \nonumber \\
\quad&+\td{A}^{n+1}\td{\theta}^{n+1}
[\sigma_2^n A_{\gamma}(2\Delta t,\td{T}_2^{n+1})
+\sigma_2^{n+1} B_{\gamma}(2\Delta t,\td{T}_2^{n+1})] I_{1,2}(\td{T}_2^{n+1})
\nonumber \\
\quad&-\td{A}^{n+1}\td{\theta}^{n+1}
\sigma_2^n A_{\gamma}(2\Delta t,\td{T}_1^{n+1})
+\sigma_2^{n+1} B_{\gamma}(2\Delta t,\td{T}_1^{n+1})] I_{2,2}(\td{T}_i^{n+1})
\notag \\
\notag \\
&A_{\gamma}(\Delta t,X) = \left[-\exp(-2\Delta t/X)+
 \frac{1-\exp(-2\Delta t/X)} {2\Delta t/X}\right]
\frac{1}{1-\exp(-2\Delta t/X)}	\notag \\
&B_{\gamma}(\Delta t,X) = \left[1-\frac{1-\exp(-2\Delta t/X)}{2\Delta
  t/X}\right]\frac{1}{1-\exp(-2\Delta t/X)} 	\notag 
\end{align}
\hrule
\label{tab:sch2c}
\end{table}

The stochastic integrals $\td{\gamma_1}^{n+1}$ and $\td{\gamma_i}^{n+1}$ 
remains to be defined.
Their expression are computed using the same hypothesis,
{\it i.e.} we suppose that
\begin{equation}
\label{eq:si_int}
\sigma_i(s,\mb{x}(s)) = \sigma_i(t_0,\mb{x}(t_0)) + \frac{1}{\Delta t}
[\sigma_i(s,\mb{x}(s))-\sigma_i(t_0,\mb{x}(t_0))](s-t_0).
\end{equation} 
After the integration by parts has been carried out,
all the stochastic integrals are manipulated in order 
to express them as functions of the base integrals 
(\ref{I1i})-(\ref{I3i}), as done in section \ref{sec:sol}.
Then, final expressions are obtained
\begin{eqnarray}
\gamma_i^{n+1} &=& [\sigma_i^n A_{\gamma}(2\Delta t,\td{T}_i^{n+1}) +
\sigma_i^{n+1} B_{\gamma}(2\Delta t,\td{T}_i^{n+1})] I_{1,i}(\td{T}_i^{n+1}) 
\quad \text{i=\{2,3\}}\\
\gamma_1^{n+1} &=& [\sigma_1^n A_{\gamma}(2\Delta t,\td{T}_1^{n+1}) +
\sigma_1^{n+1} B_{\gamma}(2\Delta t,\td{T}_1^{n+1})] I_{1,1}(\td{T}_i^{n+1}) 
\quad \text{i=1} \nonumber \\
&+&\td{A}^{n+1}\td{\theta}^{n+1}
[\sigma_2^n A_{\gamma}(2\Delta t,\td{T}_2^{n+1})
+\sigma_2^{n+1} B_{\gamma}(2\Delta t,\td{T}_2^{n+1})] I_{1,2}(\td{T}_2^{n+1})
\nonumber \\
&-&\td{A}^{n+1}\td{\theta}^{n+1}
\sigma_2^n A_{\gamma}(2\Delta t,\td{T}_1^{n+1})
+\sigma_2^{n+1} B_{\gamma}(2\Delta t,\td{T}_1^{n+1})] I_{2,2}(\td{T}_i^{n+1})
\end{eqnarray}
where the functions $A_{\gamma}(\Delta t,X)$ and 
$B_{\gamma}(\Delta t,X)$ are given by
\begin{equation} 
\begin{split}
& A_{\gamma}(\Delta t,X) = \left[-\exp(-2\Delta t/X)+
 \frac{1-\exp(-2\Delta t/X)} {2\Delta t/X}\right]
\frac{1}{1-\exp(-2\Delta t/X)}	\\
& B_{\gamma}(\Delta t,X) = \left[1-\frac{1-\exp(-2\Delta t/X)}{2\Delta
  t/X}\right]\frac{1}{1-\exp(-2\Delta t/X)}. 
\end{split}
\end{equation}
It is essential that the stochastic integrals are simulated by 
the same ${\mc N}(0,1)$ random variable used in the simulation of
the stochastic integrals $I_{i,j}^n$ computed in the Euler scheme.
In appendix \ref{app:order} it is shown
by stochastic Taylor expansion \cite{Klo_92}  that the
present scheme is a weak order scheme of order $2$ in time for system
(\ref{eq:sysEDS}). It is worth emphasizing that no correction is
done on position, ${\mb x}$, since the prediction is already of order
$2$. This property is in line with the constraint stated in Section
\ref{sec:cont}, that is, the numerical scheme should minimize the
procedures where particles must be located in a mesh (which is done
every time the particles are moved, \textit{i.e.} when a new value of
$\mb{x}$ is computed). The complete scheme is summarized in Table
\ref{tab:sch2c}.

\subsubsection{Limit cases} \label{sec:lim_sch2}
When the flow becomes laminar, that is when $T_i \to 0$ with no
condition on the product $\sigma_{i}\, T_i$, one has the following
limits: $A(\Delta t,T_i) \to 0$, $B(\Delta t,T_i) \to 1$ and
$\gamma_i(t) \to 0$, which gives for the fluid velocity,
\begin{equation} \notag
U_{i}^{n+1}=\lra{U_{i}^{n+1}}.
\end{equation}
It can rapidly be shown, by regular Taylor expansion, that this
scheme, together with the prediction step (Euler scheme sch1) is a
second order scheme for system (\ref{eq:limc1}).
In the other limit,the fluid velocity 
become a fast variable (limit case (ii)), that is when
$1 \ll \Delta t/T_i$ and
$\sigma_{i}\,T_i \rightarrow cst$, one can write 
\begin{equation} \label{eq:lim_uti1}
U_{i}^{n+1} = \lra{U_{i}^{n+1}} +
\sqrt{\frac{[\td{\sigma_i}]^2\,\td{T_i}^{n+1}}{2}}\;{\mc G}_{1,i},
\end{equation}
which is in line with the previous results.
In limit case (ii), the numerical scheme
for the position of the  particles is equivalent to the Euler scheme
written previously and is of first order in time. 

\subsection{Consistency and order of the scheme} \label{app:order}
In this section, we show by stochastic Taylor expansion that the numerical scheme
proposed is weak second-order.
For a general set of SDEs
\begin{equation}\notag
dy_i = D_i({\bf y})y_i + \sigma_{ij}({\bf y})dW_j
\end{equation}
the general condition to be fulfilled 
is that \cite{Klo_92}
\begin{eqnarray}\notag
y_i^{n+1} &=& y_i^n + D_i^n\Delta t + \sigma_{ij}^n \Delta W_j \\
&+&\frac{1}{2}\left(\frac{\partial D_i^n}{\partial y_k}\right)
\sigma_{kj}^n\Delta t \Delta W_j
+\frac{1}{2}\left(\frac{\partial \sigma_{ij}^n}{\partial y_k}\right)
D_k^n\Delta t \Delta W_j \\
&+&\frac{1}{2}\left(\frac{\partial D_i^n}{\partial y_k}\right)D_k^n\Delta t^2
+\frac{1}{4}\left(\frac{\partial^2 D_i^n}{\partial y_k\partial y_l}\right) 
\sigma_{km}^n\sigma_{lm}^n\Delta t^2 \\
&+&\frac{1}{4}\left(\frac{\partial^2 \sigma_{ij}^n}{\partial y_k\partial y_l}\right) 
\sigma_{km}^n\sigma_{lm}^n\Delta t \Delta W_j \\
&+& \left(\frac{\partial \sigma_{ij}^n}{\partial y_k}\right)\sigma_{kl}^n\int_0^tW_ldW_j
\end{eqnarray}

In present case we have
\begin{equation}\notag
D_i = \beta_{ij}y_j= 
\begin{pmatrix} 
& 0&  \dots   &   &1  &  & \\
&  & 0 &   &  & 1   &\\
&  &   & 0 &  0&   &1\\
& 0 & \dots & & -\alpha_1 & A_{12} & 0 \\
&  & 0 &  &0 &-\alpha_2 &0 \\
&  &  &0  &0  &0 &-\alpha_3 
\end{pmatrix}
\begin{pmatrix} x_1 \\
x_2\\
x_3 \\
U_1 \\
U_2 \\
U_3 \end{pmatrix}
\end{equation}
with $\alpha_i = 1/T_i$, thus
\begin{equation}\notag
\left\{ \begin{split} 
 \left(\frac{\partial D_i}{\partial y_k}\right)& = \beta_{ij}\delta_{jk}
+ \left(\frac{\partial \beta_{ij}}{\partial y_k}\right)u_j  \\
\left(\frac{\partial^2 D_i}{\partial y_k\partial y_l}\right)  & =
\left(\frac{\partial^2 \beta_{ij} }{\partial y_k\partial y_l}\right) u_j
+\left(\frac{\partial \beta_{ij}}{\partial y_l}\right)\delta_{jk}
+\left(\frac{\partial \beta_{ij}}{\partial y_k}\right)\delta_{jl}
\end{split} \right.
\end{equation}

In the equations given in tables \ref{tab:sch1}-\ref{tab:sch2c},
we proceed to Taylor developments of all deterministic functions.
Moreover, those quantities which are marked by a tilde in the correction step
are explicitly written and, then, developed.
In this way, all functions are computed at time $n$, 
and $\delta y=y^{n+1}-y^n$, thus, for the sake 
of simplicity, the upper-script is canceled from now on.
We obtain for the prediction step
\begin{eqnarray}\notag
\delta \td{x}_i &=& U_idt + O(\Delta t^{3/2}) \\
\delta \td{U}_1 &=& (-\alpha_1y_4)\Delta t + \sigma_{11}\Delta W_1 \notag \\
&-&\frac{1}{2}(\alpha_1\sigma_{11})\Delta t\Delta W_1 + 
\frac{1}{2}(A_{12}\sigma_{22})\Delta t\Delta W_2 \notag \\
&+&\frac{1}{2}(\alpha_1^2)\Delta t^2-\frac{1}{4}(\alpha_1+\alpha_2)A_{12}\Delta t^2\\
\notag \\
\delta \td{U}_i &=& (-\alpha_iy_{i+3})\Delta t +\sigma_{ii}\Delta W_i \notag \\
&+&\frac{1}{2}(\alpha_i^2)\Delta t^2-\frac{1}{2}(\alpha_i\sigma_{ii})\Delta t\Delta W_i
\;\text{with}\;i=\{2,3\}~.
\end{eqnarray}
Then, for  the correction step we use the properties of matrix $\sigma$ 
which is diagonal ($\sigma_{ij}=\sigma_i\delta_{ij}$)
 and respect the property (\ref{sig-prop}).
For the sake of simplicity we use the tensor notation for partial derivatives.

The stochastic integrals $\td{\gamma}^{n+1}_i$ give
\begin{eqnarray}\notag
\td{\gamma}_i^{n+1} &=& \sigma_i\Delta W_i\;(O\Delta t) \\+
&+&\frac{1}{2}(\partial_k\sigma_i)\sigma_k\Delta W_i\Delta W_k\;(=0)\\
&+&\frac{1}{2}(\partial_k\sigma_i)\beta_{kl}y_l\Delta W_i\Delta t+
\frac{1}{2}(\partial_j\beta_{ij})\Delta W_j\Delta t\;(O\Delta t^{3/2})\\
&+&\frac{1}{2}(\partial_k\beta_{ij})\sigma_i\sigma_k\Delta W_k\Delta W_j\Delta t
\;(O\Delta t^2)\\
&+&\frac{1}{12}(\partial_k\sigma_j)\beta_{ij}\sigma_k\delta_{kj}\Delta t^{2}\;(=0)
\end{eqnarray}
finally, we globally obtain
\begin{eqnarray}\notag
y^{n+1}_i&=&y_i^n+D_i\Delta t+\sigma_i\Delta W_i \\
&+&\frac{1}{2}\Delta t\Delta W_i\left[\partial_k\sigma_i)\beta_{kl}y_l+
(\partial_k\beta_{il})y_l\sigma_k+\beta_{ij}\sigma_j\right]\\
&+&\frac{1}{2}\Delta t^{2}\left[\beta_{ij}^2y_j
+(\partial_k\beta_{il})\beta_{km}y_ly_m
+(\partial_k\beta_{ik})\sigma_k^2
+\frac{1}{2}(\partial^2_{kl}\beta_{ij})\sigma_k\sigma_ly_j\right]
\end{eqnarray}
In the present case, the last expression coincides exactly with 
the equation required by Taylor expansion in the general case and,
therefore, it is demonstrated that the numerical scheme proposed 
is of order two in the weak sense.

In summary, a weak-second order scheme for system (\ref{eq:sysEDS})
has been derived. This scheme satisfies all conditions listed in
Section \ref{sec:cont}, except for the second order convergence
condition in limit cases (ii).
This imperfection is bearable since this case has mainly a formal importance
and it does not occur very often in practice,
moreover, it is inherent to the spirit of the scheme, that is a single
step to compute position ${\mb x}$ (in order to minimize the number
of particle localizations in the algorithm).

Finally, we want to recall which are the principal hypothesis and properties used
and what steps have been followed to achieve this weak second-order numerical scheme.
\begin{enumerate}
\item[a)] The analytical solution to the system (\ref{eq:sysEDS}) is
calculate,  taking constant the coefficient. 
It is worth emphasizing that it always possible (almost formally)
to derive analytical solutions for linear SDEs with constant coefficients
\cite{Ott_96}.
\item[b)] The complete covariance matrix of all stochastic integrals is calculated.
In the present case, the stochastic integrals have been preliminary decomposed 
on a base of simple stochastic integrals and, then, the covariance matrix
of these last ones is computed. The original integrals appearing in the equations
can be easily computed through a matrix equation.
Generally speaking, this operation is not necessary, but sometime 
it can be a powerful method to simplify involved calculations.
\item[c)]The analytical solutions are discretised to obtain 
an Euler weak first-order scheme. 
Since this scheme is derived directly from the analytical solutions,
the consistency of the scheme, even in asymptotic limits, is naturally assured.
Moreover, the presence of exponential terms does guarantee 
the global unconditionally stability, though the scheme is explicit.
\item[d)] Thanks to the particular structure of the diffusion matrix $\sigma$,
Eq. (\ref{sig-prop}), it is possible to chose
a predictor-corrector algorithm,
with correction only on fluid velocity,
 in order to develop a weak second-order scheme.
The first-order scheme is taken as prediction step.
\item[e)] The correction is developed starting from analytical solutions
of fluid velocity and letting all coefficients to vary linearly in time-step
(hypothesis of constant gradient).
The basic idea is that the variation of higher order ( linear gradient etc)
can contribute with terms at least of order 3 in time-step,
as all coefficients appear  inside integrals.
\end{enumerate}
It is worth underlying, in conclusion at this section,
that the numerical scheme proposed remains
of weak second-order even in the high-Reynolds number flows, 
that is for $\nu = 0$, when the model becomes the well known
SLM model \cite{Pop_94}.
Furthermore, it ever fills all the constraints 
asked in section \ref{sec:cont}.
\section{Boundary conditions} \label{sec:BC}
In the present Lagrangian approach the no-slip and impermeability conditions
are satisfied by imposing the boundary conditions on stochastic particles.
A stochastic particle is assumed to have crossed the wall boundary 
(placed at $y=0$ if 
at the end of time-step $Y(t+\Delta t)=y_{out}<0$.
For those reflected particles we impose the conditions
\begin{equation}
y_{in} = |y_{out}| \; ; \; \text{and} \; {\bf U}(t+\Delta t)=0~.
\end{equation}
Symmetry conditions are imposed at the other boundary placed 
a the center of the channel $y=h/2$. 
Along the $x$ direction, periodic conditions are imposed.

The problem of boundary conditions for walls
in stochastic method remains a fascinating and challenging open issue.

\section{Numerical Results}
This section present some numerical results which complement numerical simulations discussed 
elsewhere~\cite{Chi_09ijmf}.

A channel-flow
is solved both with an Eulerian method and, then, with the PDF one.
Being a particle-mesh method, particles are moving in a mesh where, at every
cell, the mean fields describing the fluid are known. 
Generally speaking, the statistics
extracted from the variables attached to the particles, which are
needed to compute the coefficients of system (\ref{modelp})-(\ref{modelv})
are
not calculated for each particle (this would cost too much CPU time)
but are evaluated at each cell center following a given numerical
scheme (averaging operator). These moments can then be evaluated for
each particle by projection. This the general principle of particle-mesh
methods: exchange data between particle and mesh points. Once again,
the expected value for functionals of the state vector are not
computed directly for each particle but are evaluated at discrete
points in space and then calculated for
each particle by interpolation.
The averaging and projecting operators are computed by 
near grid point (NGP) method \cite{Hoc_88}.
In our particular case, the same physical case 
is considered by both methods,
thus, first the flow is computed by the moment method, then, 
the computed mean fields are used, frozen in time, in the simulation 
of stochastic particles.
Therefore, there is not two-way coupling between the two methods 
and issues concerning averaging and projection of variables are not 
much influent, provided that an interpolation enough refined is used.
It is 
indeed important to note that the interpolation and backward estimation schemes introduce some errors~\cite{Gar_07,Gar_09}. Nevertheless, for the present simulations, the very fine mesh 
used in the DNS (128 points along $y$) is also used here, which should assure a good 
interpolation of the mean fields to the Lagrangian stochastic particles and make these 
errors negligible in this case. 
Furthermore, it is worth saying that results obtained in the deterministic context 
do not generally apply to the stochastic one~\cite{Pei_06}.
\subsection{Consistency}\label{sec:cons}
\vspace{0.2cm}
\begin{figure}[h!]
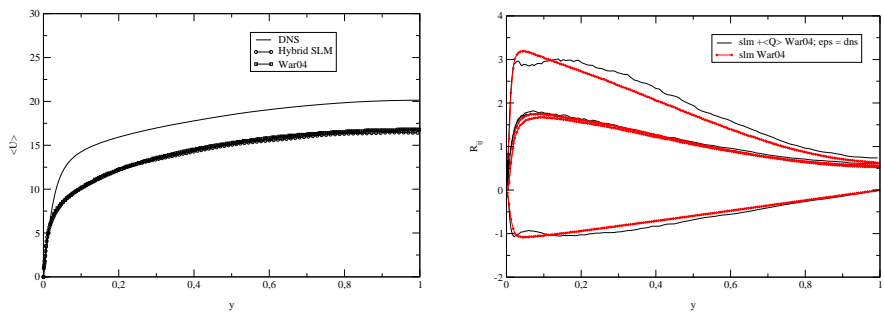

\begin{center}
{\epsfig {file=Figure/medie.eps,width=5.5cm}}
\hspace{0.5cm}{\epsfig {file=Figure/rey-Pozall-e-dns2.eps,width=5.5cm}}
\caption{(a) Mean velocity Results in Wacawczyk {\it et al.} \cite{Wac_04} (War04), in the present Hybrid configuration
and in the DNS. The results are in non-dimensional units. (b) Reynolds stress results in War04 and in present Hybrid configuration,
the results are in non-dimensional units.}
\label{Fig_1}
\end{center}
\end{figure}
The computations have been performed for the case of fully developed turbulent
channel flow at $Re = u_{\tau}h/\nu = 395$.
The channel flow DNS results of Moser {\it et al.} \cite{Mos_99}
have been taken for comparison as reference data.
For the simulations, 
the mesh used in DNS (128 points along $y$) is also used here,
which assures a good interpolation of the mean fields 
for the Lagrangian stochastic particles.
Quantities designed  with the upper-script $+$
are non-dimensionalised with wall parameters.
Given that the PDF used is consistent with the Eulerian Rotta Reynolds-stress model
(see appendix \ref{app:reystress}), this model should be used in the Eulerian 
part of the Hybrid method in order to check the consistency of the Lagrangian part.
To ensure the best possible consistency,
 we  use the mean fields  obtained 
with an equivalent Lagrangian model ~\cite{Wac_04} 
in a stand-alone configuration.
In practice, the Eulerian mean-fields computed 
from the Lagrangian stand-alone simulations carried on by 
Wacawczyk {\it et al.} \cite{Wac_04} (designed as War04)
are employed, here, as Eulerian counterpart.
Indeed, the equations for the moments of first and the second order
are exactly the same for both models and, thus, a perfect consistency 
is assured.
Nevertheless, in that stand-alone calculation
turbulent dissipation $\epsilon$ is not computed directly,
therefore,
for this variable DNS values are used and, thus,
this remains a possible source of inconsistency. 

In figures \ref{Fig_1}, three different profiles for the mean-velocity 
are shown:
the profiles obtained in the stand-alone configuration in the paper
of Wacawczyk {\it et al.} with an analogous viscous model,
 the profiles obtained in the hybrid configuration  and 
the DNS profile, for physical reference.
In figures \ref{Fig_1}, the profiles of second-order moments,
that is the Reynolds stress tensor,
are shown, for the present and for the War04 calculations.
Generally speaking, since both methods face the same test-case,
if they were completely consistent,
the results would be equal both for first and
second-order moments.
The profiles deriving from the present PDF method are in good agreement
with those computed in the stand-alone configuration.
Some small differences are normal for the different numerical framework
and since a different profile is used for $\epsilon$ .

Yet, in figures \ref{Fig_1},
it is possible to observe that in the zone
of $y ~\sim 0.05-0.1 (y^+ ~\sim 10-50)$
the first components of Reynolds stress $R_{11}$ 
attain a nearly constant level.
This merits a particular attention.
\vspace{0.5cm}
\begin{figure}[h]
\begin{center}
{\epsfig {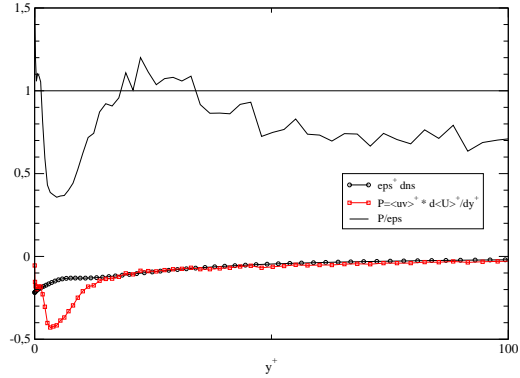}}
\caption{Comparison between dissipation and production.}
\label{Fig_2}
\end{center}
\end{figure}
This behavior can be traced back to the value of turbulent dissipation
$\epsilon$ used in the present calculations. 
In figure \ref{Fig_2}, the dissipation and production 
 profiles are shown as well as  their ratio.
It is clear that in the zone of interest a relation of equilibrium
exists,  the value of the ratio is near to 1 and, then, 
it becomes of  about $0.8$ till $y^+ ~\sim 100$.
In this range, the Reynolds-stress equations lead to the relations
\begin{equation}
\lra{u^2}=k\frac{C_0+2{\mc P}/\epsilon}{1+3C_0/2},\;\; 
k=u_{\tau}^2\frac{1+3C_0/2}{\sqrt{C_0}}=u_{\tau}^2k_{cst}~.
\end{equation}
and, then, using the hypothesis ${\mc P}/\epsilon=1$
\begin{equation}
\lra{u^2}=u_{\tau}^2\frac{C_0+2}{\sqrt{C_0}}~,
\end{equation}
therefore, it is correct to obtain constant value of the first Reynolds stress 
components in the range $(y^+ ~\sim 10-50)$, in the present configuration.
\begin{figure}[!h]
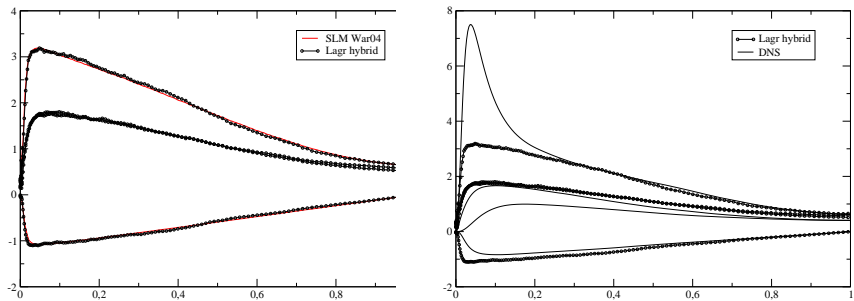

\begin{center}
{\epsfig {file=Figure/Val-fin.eps,scale=0.25}}
\hspace{0.2cm}{\epsfig {file=Figure/Val-fin-DNS.eps,scale=0.25}}
\caption{(a) Reynolds stress, final comparison. (b) Present results in the hybrid configurationj and DNS.}
\label{Fig_3}
\end{center}
\end{figure}
To complete our numerical validation about the consistency 
of the method, a slight ad-hoc modification of $\epsilon$ has been worked out.
In practice, the relation between production and dissipation has been 
imposed to be $0.9$ in the region 
$y^+ ~\sim 10-50$, while in the rest of the domain 
the DNS values are maintained.
In figure \ref{Fig_3}, the results are shown.
The two profiles overlap. 
That emphasizes the consistency of the numerical method used,
since the features of the Lagrangian model are perfectly 
reproduced.
We recall that the ingredients necessary to reach this objective
are:
\begin{enumerate}
\item[(i) ] Consistent physical model.
\item[(ii)] Consistent numerical scheme
\item[(iii)] Accurate global numerical method , 
concerning also the exchange of information from Eulerian solver
to Lagrangian one.
\end{enumerate}
Given that the method is found to be consistent,
present calculations are also compared with DNS profiles
in figure \ref{Fig_1} and \ref{Fig_3}.
From a physical point of view, the results are coherent with
model features. 
Indeed, it is well known that the SLM model, 
like the Rotta RSM model, underestimates greatly the value of the 
components $R_{11}$ and is rather isotropic in the other 
diagonal components \cite{Pop_97,Wac_04}.
 
\subsection{Asymptotic limits} \label{sec:limnum}
\begin{figure}[h]
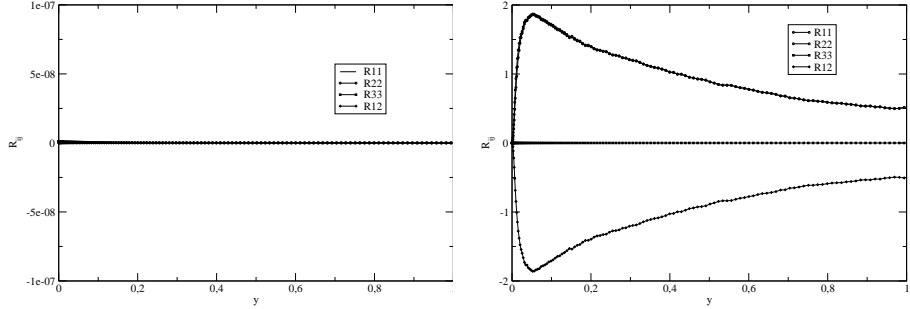

\begin{center}
{\epsfig {file=Figure/limcase1.eps,scale=0.25}}
\hspace{0.2cm}{\epsfig {file=Figure/limcase2.eps,scale=0.25}}
\caption{(a)Laminar limit case. (b) Theoretical asymptotic limit case.}
\label{Fig_4}
\end{center}
\end{figure}
A study of the numerical method in some asymptotic cases
is necessary to consider the numerical validation achieved.
Two cases are chosen, (a) the laminar limit case, essential
in wall-bounded flows for the presence of the viscous sub-layer,
and (b) a academic case where the matrix ${\bf A}$ goes to infinity,
thus the time-scales $T_1$ and $T_3$ are taken to be zero,
and $T_2=T_L$.

From a theoretical point of view, in these limit cases
the expected behavior can be anticipated.
Starting from Reynolds-stress equations, by using all hypothesis 
valid for the channel flow and considering the stationary case,
the Reynolds-stress equations can be simplified.
The equations for first, second diagonal components 
and for the shear stress are
\begin{eqnarray} 
\frac{\partial \lra{v^3}}{\partial y} &= &2G_{22}\lra{v^2}+\sigma_2^2, 
\,\,\,\quad\text{with}\,G_{22}=-1/T_L , \quad
\Rightarrow \; \lra{v^2}\approx\frac{T_L}{2}\sigma_2^2~, \label{rey_asy} \\
\frac{\partial \lra{u^2v}}{\partial y} &= &-2\lra{uv}\frac{\partial \lra{U}}{\partial y}
+2G_{11}\lra{u^2}+2A_{12}\lra{uv}, 
\,\,\,\quad\text{with}\,G_{11}=-1/T_1  \label{rey_asy2}\\
\frac{\partial \lra{uv^2}}{\partial y} &= &-2\lra{v^2}\frac{\partial \lra{U}}{\partial y}
+2G_{11}\lra{uv}+2G_{22}\lra{uv}+2A_{12}\lra{v^2} \label{rey_asy3}
\end{eqnarray}
thus, as expected, in the laminar case (a) we have to find 
\begin{equation}
T_i\rightarrow0  \;\Rightarrow \;R_{ij}=0~.
\end{equation}
In the other case(b)  we have
\begin{equation}
A_{11}=A_{33}=A_{12}\longrightarrow \infty 
\;\Rightarrow \;T_1=-\frac{1}{A_{11}}\;=\;T_3=-\frac{1}{A_{33}}\longrightarrow 0
\end{equation}
hence, from the equations (\ref{rey_asy2})-(\ref{rey_asy3})
we deduce 
\begin{equation}
\lra{u^2}=-\lra{uv}\;,\; 
\lra{uv}=-\lra{v^2}=-\frac{T_L}{2}\sigma_2^2~.
\end{equation}

The numerical results in figure \ref{Fig_4} show that
the asymptotic limits are perfectly recovered without changing time-step.
\vspace{1cm}
\begin{figure}[h]
\begin{center}
\label{Fig_5}
\end{center}
\end{figure}

\subsection{Hybrid-method error}

In the last section we have studied and verified the consinstency 
of our hybrid method. In particular, we have checked the behaviour of the method
in a theoretically consistent configuration and in the asymtotic limits.
Since the consistency of the method has been successfully proofed,
we can now study the global error which is eventually introduced 
by using an hybrid method not completely consistent.
In order to better explain this very important point,
let us start by considering the two phases separately.

The hybrid method for fluid-particle system exploites two different 
numerical (and theoretical) approaches.
For  fluid phase, a classical moment approach is used.
This kind of algorythm is characterised by the deterministic 
discretisation error that is related to the numerical integration
of equations in time and space. For particle phase, a Monte Carlo method is used to simulate the PDF
of the stochastic process which describe particle dynamics.
This method is affected both  by  deterministic discretization errors and, further,
by the Monte Carlo error, which is related to the finite
number of stochastic particles used in numerical simulations.
In the stand-alone approach to stochastic equations for turbulent flows,
the mean quantities present in  models of  form (\ref{eq:MK}) 
are computed starting from the same stochastic particles.
This procedure causes a typical Bias error 
which is found to be dominating  in stand-alone methods
 \cite{Xu_99}.
In hybrid methods, this error is avoided by using a moment approach 
to compute the mean quantities that are necessary.
In fact, this is the very strenghtness of the method.
Nevertheless, even in hybrid configurations the introduction 
of variables computed externally may be source of errors.
In particular, it is not obvious {\it a priori}  what happens
when the mean quantities are computed by a method which 
is not completely consistent with the PDF one.
The error introduced by this inconsistency is 
intrisically inherent to the kind of hybrid method used.
Moreover, from a practical point of view, the best 
way to investigate this effect is  in a hybrid fluid-fluid configurations,
where the influence of mean variables can be easily traced back.

To isolate the hybrid error, all the other errors have been 
made negligible in the following simulations.
In both methods, a time-step of $10^{-4}$ 
and a spatial-step of $10^{-4}$ have been used, 
which can be considered as zero for our purposes.
$5*10^{4}$ particles have been employed,
which can be considered infinity.
To the sake of clarity, the configuration discussed in last section,
where stand-alone  mean variables were used, it will be called 
{\it standard} configuration in the following.
\begin{figure}[h!]
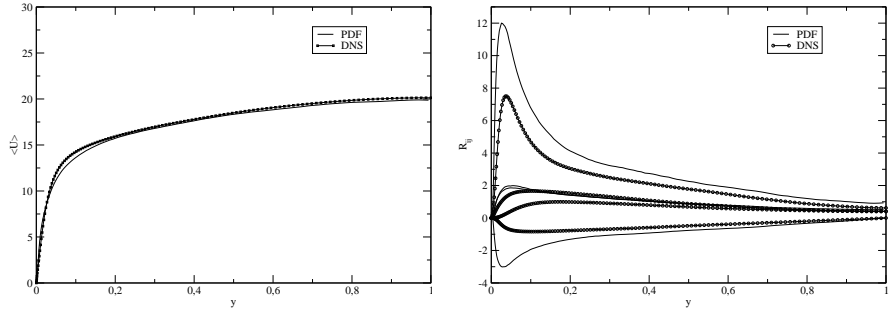

\begin{center}
{\epsfig {file=Figure/rey_gp1.eps,scale=0.25}}
\hspace{0.2cm}{\epsfig {file=Figure/rey_gp2.eps,scale=0.25}}
\caption{(a) DNS-PDF configuration: mean velocity. (b) DNS-PDF configuration: Reynolds stress.}
\label{Fig_6}
\end{center}
\end{figure}

First, we consider the following configuration:
DNS results are used to provided the PDF solver 
with all necessary mean quantities.
In figure \ref{Fig_6}, we present the mean velocity computed by the PDF method
in this configuration.
The DNS mean velocity is also shown as well as the mean velocity
computed by PDF method in the consistent PDF-PDF configuration.
The mean velocity so obtained is perfectly in agreement 
with DNS one and, thus, is strongly different from that obtained
in the {\it standard} configuration.
In fact, the exact profile is now recovered.
In figure \ref{Fig_6}b, the Reynolds stress are shown for the same 
 configuration.
The profiles are dramatically changed in comparison with {\it standard} results.
The $\lra{uv}$  and, as a consequence as we have seen,
 the $\lra{u^2}$ components are strongly overpredicted now.
On the contrary, the other diagonal components are not changed.

This behaviour can be explained, at least approxiamatively.
We have seen in the last section, 
analysing the behaviour in the asymptotic limits,
that the diagonal $\lra{v^2}$ and $\lra{w^2}$ components are essentially
dependent on Lagrangian time-scale and on diffusion coefficient,
which have not  changed and, therefore, they remain the same.
The cross shear $\lra{uv}$ depends on the gradient of the mean velocity,
other than on turbulent kinetic energy and on Lagrangian time-scale,
thus it is much sensisitive to changes in mean velocity and, in fact,
 the effect of this is enourmous.
Indeed, the present mean velocity profile 
atteins a higher maximum than in the {\it standard} configuration and,
thus, it is much steeper.
The shape of  $\lra{u^2}$ is a direct consequence of this behaviour.
\vspace{0.5cm}
\begin{figure}[!h]
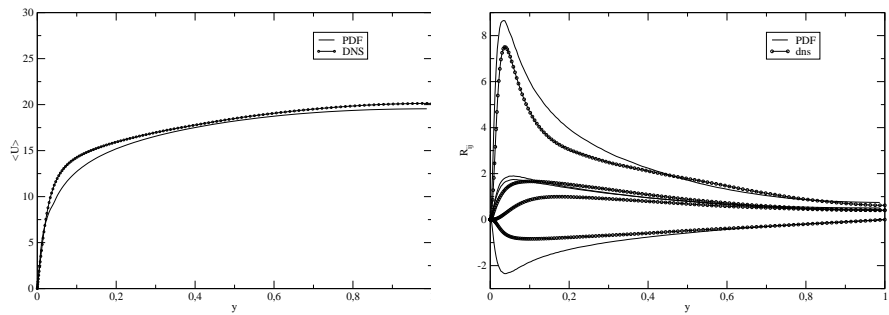

\begin{center}
{\epsfig {file=Figure/v2f-dns-U.eps,scale=0.25}}
\hspace{0.2cm}{\epsfig {file=Figure/v2f-dns-R.eps,scale=0.25}}
\caption{(a) V2F-PDF configuration: Mean velocity. (b) V2F-PDF configuration: Reynolds stress.}
\label{Fig_8}
\end{center}
\end{figure}

Now, we change again our configuration and we couple 
the PDF solver to a low-Reynolds number RANS model,
known for his good performance in boundary flows,
the v2f model \cite{Dur_91}.
In the present calculations, we use a refinement version 
of the model \cite{Lau_05}.
In figure \ref{Fig_8}, we show the mean velocity profile 
togheter with DNS and standard PDF results.
As previously, the mean velocity given by our PDF method is 
quite near to the Eulerian mean provided to the PDF solver,
in this case computed by v2f solver.
Furthermore, this result is also in good agreement 
with DNS result.
In figure \ref{Fig_8}, Reynolds stress profiles are shown.
It can be seen that results are quite similar to those obtained 
in the DNS-PDF configuration.
This time, the $<uv>$ and the  $<u^2>$ components 
are  less overpredicted  but they show qualitatively
the same behaviour.

This series of results leads us to some conclusions.
\begin{itemize}
\item[i)] In all three configurations tried, the mean velocity
computed by stochastic particles collapses on the value
of the Eulerian mean velocity that is provided by the Eulerian solver.
This is in line with the physics of the model,
which is basically base on a return-to-equilibrium idea \cite{Min_97}.
\item[ii)] In all cases, but the standard one, there is a difference between Eulerian and Lagrangian 
results at the level of second-order statistical moments (Reynolds stress).
This is a conseguence of coupling in a Hybrid approach 
two methods which are not consistent,
this reasoning is valid both for PDF-DNS and for PDF-V2F configurations.
Thus, it is possible to identify the global error 
of hybrid Eulerian/Lagrangian method by comparing
the actual Reynolds stress profiles with those computed
in the complete consistent configuration \ref{Fig_3}.
\item[iii)] In the last two configurations tried here,
the results obtained  are quite similar in practice, 
even though DNS and V2F approaches are quite different from a theoretical point of view.
This is reasonable.
In fact, the  profiles provided by the Eulerian solver to the Lagrangian one
are mean velocity, mean pressure, turbulent energy and turbulent dissipation.
For these variables, the V2F approach gives results in very good agreement with
DNS ones. Therefore, from the point of view of the hybrid method
DNS and V2F approaches are very near.
Nevertheless, the results which have been obtained using V2F are nearer to 
those consistent than those obtained using DNS, figure \ref{Fig_6}.
Therefore, V2F model is found to be more consistent with present Lagrangian model 
than DNS, as expected.
\end{itemize}

The large error in the off-diagonal component of the Reynolds stress $\lra{uv}$
might cause some doubts about the results shown in this section.
It is possible to verify by a simple numerical trick that
these results could be expected.

\begin{figure}[ht]
\begin{center}
{\epsfig {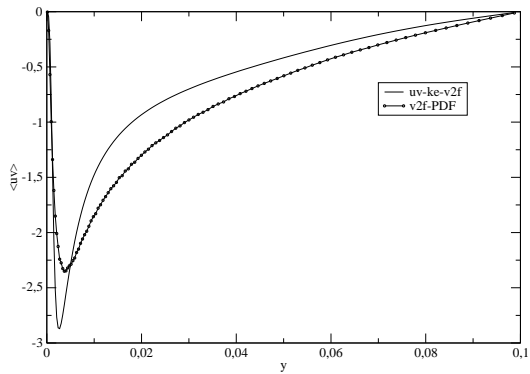}}
\caption{Reynolds stress from V2F results through $k-\epsilon$ formula.}
\label{Fig_10}
\end{center}
\end{figure}

Starting from the mean profiles obtained by the V2F approach,
we can compute the $\lra{uv}$ component of Reynolds stress by using
the $k-\epsilon$ formula
\begin{equation}
\lra{uv}=C_{\mu}\frac{k^2}{\epsilon}\frac{\partial \lra{U}}{\partial y}~.
\end{equation}
The resulting profile is shown in figure \ref{Fig_10} and it is compared
with that obtained in the hybrid PDF-V2F configuration for the same variable.
The profiles are qualitatively similar. 
This shows that the PDF method used is approximatetly consistent 
with the $k-\epsilon$, 
even though it is rigorously consistent with the Rotta Reynolds stress model.
Furthermore, it demostrates that the difference  between 
the V2f model and the $k-\epsilon$ model 
introduces an error which generates the behavior encountered for $\lra{uv}$.

\subsection{Standard model Vs Viscous model}

In this work we have proposed and tested a new Langevin viscous model 
for turbulent fluid flows.
Other propositions have been recently made,
with the aim of improving the present description 
of near-wall layer in the framework of PDF approach.
In this section, we compare the viscous model 
with the standard high-Reynolds number one, in the hybrid simulation
of turbulent channel flow.
The high-Reynolds number model is generally used with wall-function
boundary conditions, without integrating up to the wall.
In the following computations, we proceed to the up-to-the-wall integration
and we use the same boundary conditions imosed in the viscous case \ref{sec:BC}.

\begin{figure}[ht]
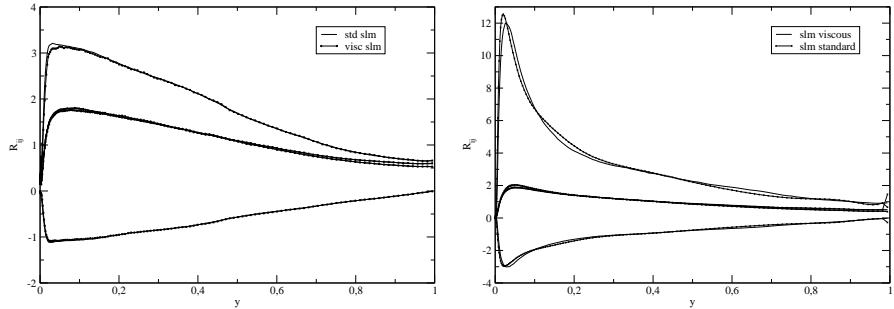

\begin{center}
{\epsfig {file=Figure/visc-vs-std1.eps,scale=0.25}}
\hspace{0.2cm}{\epsfig {file=Figure/visc-vs-std2.eps,scale=0.25}}
\caption{(a) Reynolds stress for standard and viscous model, in the consistent configuration. (b) Reynolds stress for standard and viscous model, 
in the DNS/PDF configuration.}
\label{Fig_11}
\end{center}
\end{figure}
First, we carry out the simulation of the channel flow in the standard
consistent configuration (section \ref{sec:cons}).
In figure \ref{Fig_11}, the Reynolds stress obtained in both cases 
are shown. The differences are  not important, in practice, 
the two models give the same results.
In figure \ref{Fig_11}, the Reynolds stress obtained in the 
hybrid DNS/PDF configuration are shown.
Also in this inconsistent configuration, the results obtained
with standard and viscous model do not show significant differences.

\begin{figure}[h]
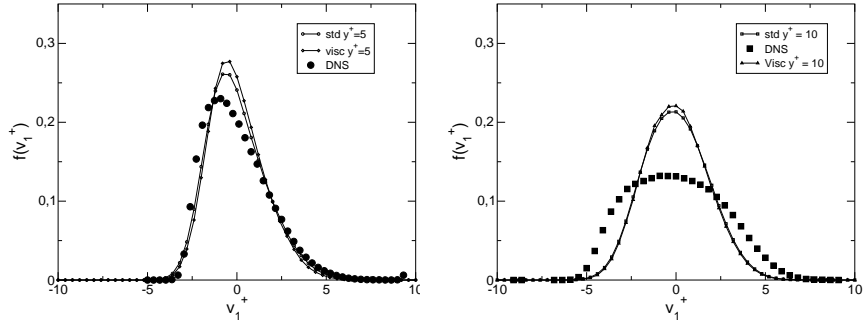

\begin{center}
{\epsfig {file=Figure/PDF5_hybrid.eps,scale=0.25}}
\hspace{0.2cm}{\epsfig {file=Figure/PDF10_hybrid.eps,scale=0.25}}
\caption{(a) Probability density functions computed by viscous, standard
PDF and DNS approach
at $y^+=5$. (b) Probability density functions computed by viscous, standard 
PDF and DNS approach
at $y^+=10$.}
\label{Fig_13}
\end{center}
\end{figure}

In figures \ref{Fig_13}, the probability density functions 
of the normal velocity are shown for two locations: $y^+=5$ and $y^+=10$.
The PDFs are computed with both standard and viscous PDF methods and, for reference,
also DNS results are shown.
Once again, the new viscous terms bring very small changes to PDF profiles.
From a general point of view, it is possible to observe that in this configuration
with the SLM model the PDFs remain quasi-Gaussian and 
the distorsion due to walls is scarcely felt, 
mainly in the transition zone at $y^+=10$.
In that sense, given the good results shown by PDF methods equipped by
elliptic relaxion model \cite{Dre_98,Poz_04}, it is possible to conclude
that in order to improve the quality of the physical picture
the essential  is given by non-local representation 
of pressure fluctuating term.

In conclusions, the viscous correction in near-wall models
are important from a theoretical point of view,
for they lead to the exact expression for first and second moment equations.
However, from a practical point of view, they are negligible
in hybrid configurations.
Indeed, the numerical results are not influenced in a noticeable way 
by the viscous corrections.
Other viscous models have been proposed \cite{Dre_98,Wac_04},
but this is the first time that such  comparison has been made.
Moreover, the other models have been studied in  stand-alone 
configurations, where the integration up to the wall is more delicate.
Thus, it is not possible to generalize this conclusion to 
all PDF methods, in stand-alone methods viscous terms may be necessary 
in order to compute directly the near-wall zone.

\section{Conclusions}

First, a new physically-consistent PDF  model is proposed, 
then, a numerical scheme for the stochastic differential equations 
which appear is detailed. 
Finally, a numerical validation of the global method
is carried out with the main attention pointed on 
the issue of the consistency between the two approaches.

The model proposed is in the form of a set of stochastic differential equations
(\ref{modelp})-(\ref{modelv}).
Since the model belongs to the same category
of those studied previously \cite{Pop_97,Wac_04},
a physical validation of the model, 
to be done necessarily with a stand-alone code, 
should not show particular new insights.
Therefore,
we suppose that the performances of the present model 
are equivalent to those already validated \cite{Wac_04}
and we have chosen to put the emphasis on the numerics.
With regard to thos previous models, the present one is consistent with the same first and second order moment equations and thus represents just an alternative.
Its main feature is that takes into account of the viscosity without changing the x-equation,
as done in previous models. This presents and interest since allows a easier treatment of boundary conditions.

To simulate the stochastic process,
 a numerical scheme is proposed and discussed into details.
This numerical scheme has been developed with some fundamental 
constraints in mind:
\begin{enumerate}
\item[(i)] \textit{The numerical scheme must be explicit, uncoditionnaly stable, of
order $2$ in time and the number of calls to particle localization has
to be minimum}.
\item[(ii)] \textit{ The numerical scheme must be consistent with the
analytical solutions of the system when the coefficients are constant}.
\item[(iii)] \textit{The numerical scheme must be consistent with all
limit systems}.
\end{enumerate}
These constraints are not mathematical strangeness but they come out
from the intrinsic structure of the stochastic system and are important for 
practical concerns.
Indeed, they assure consistency, accuracy and efficiency 
to the numerical method, even for equations which present a multi-scale character.
In particular, in bounded flows velocity time-scale goes to zero
with approaching to the wall and the stochastic system 
(\ref{eq:sysEDS}) becomes stiff. 
An algorithm which is not consistent with this asymptotic limit
and it is not uncoditionnaly stable would require a related small
integration time-step, 
making numerical simulations practically impossible.
The consistency has been demonstrated analytically.

The numerical method obtained is validated in a Hybrid 
Eulerian/Lagrangian configuration.
Usually a new numerical scheme is
validated with a study of different errors arising from the numerical scheme, 
in a stand-alone configuration.
Nevertheless, the problem has been analyzed exhaustively
in the case of free-shear fluid flows with another weak second-order scheme
\cite{Xu_99} and in the case of two-phase flows for an analogous scheme
\cite{Min_03}.
In particular, the last scheme is retrieved  rigorously by the present scheme
in the case of high Reynolds-number flows (when viscosity is put to zero).
Thus, we have preferred to concentrate ourselves on the major point 
of the consistency of the hybrid Eulerian/Lagrangian method, 
which is a fundamental issue for this kind of approach \cite{Mur_01}.
The global method is found to be perfectly consistent at the level 
of first two velocity moments.
Moreover, analytical results for Reynolds stress 
are recovered in the asymptotic limits, 
which supports also the numerical validation of the numerical scheme.

In this numerical validation, it has been pointed out two other main points: (i) Eulerian/Lagrangian methods which are not consistent with the same turbulence model like DNS/SLM or V2F/SLM suffer from an important bias error which make result flawed. Therefore such configurations often proposed 
in two-phase flows should reconsidered critically. (ii) It has been shown that in a hybrid framework the viscous and non-viscous models give basically the same results, provided the correct boundary conditions are imposed.   

Finally, the main point to underline, about the development of the numerical scheme,
is that the methodology presented goes beyond the borders of the present model
and of present calculations.
The main objective was  to propose a safe guideline
to follow, when one has to deal with numerical simulations of stochastic models.
Moreover, the methodology is not restricted to fluid mechanical 
applications, but it can be applied without modifications
to whatever model which has a form of linear SDEs,
for example in polymeric fluids \cite{Ott_96}.

\appendix
\numberwithin{equation}{section}

\section{Properties of the physical model}\label{app:reystress}

The mean
momentum equation is simply obtained by applying the averaging operator
to the particle velocity equation (\ref{modelv})
\begin{equation}
\lra{ dU_i } = -\frac{1}{\rho}
\frac{\partial \lra{ P }}{\partial x_i}\, dt
+\nu\frac{\partial^2 \lra{ U_i }}{\partial x_k^2}.
\end{equation}

Using the relation between the instantaneous substantial derivative and
its Eulerian counterpart,
$d \cdot /dt=\partial\cdot /\partial t + U_j \partial\cdot/\partial x_j$,
we obtain
\begin{equation}
\frac{\partial \lra{ U_i }}{\partial t} + 
\lra{ U_j} \frac{\partial \lra{ U_i }}{\partial x_j}
+ \frac{\partial \lra{ u_iu_j }}{\partial x_j} =
-\frac{1}{\rho}\frac{\partial \lra{ P }}{\partial x_i}
+\nu\frac{\partial^2 \lra{ U_i }}{\partial x_k^2}.
\end{equation}
Thus, the exact mean Navier-Stokes equation is
satisfied.  This should not be too surprising since convection is
treated without approximation by the Lagrangian point of view and since
the mean pressure-gradient, which represents the mean value of the
acceleration of a fluid particle, is properly taken into account in
Eq.~(\ref{modelv}).  
For the second order,   one has
to write the instantaneous equations for the fluctuating velocity
components along a particle trajectory.  This is done by writing $u_i =
U_i-\lra{ U_i}$ and consequently
\begin{equation}
\frac{du_i}{dt}=\frac{dU_i}{dt}- \left\{
\frac{\partial \lra{ U_i }}{\partial t} +
\lra{ U_j} \frac{\partial \lra{ U_i }}{\partial x_j}
\right\} - u_j\frac{\partial \lra{ U_i }}{\partial x_j}.
\end{equation}

We now write the equation in an incremental form to properly handle
the stochastic terms, which using the mean Navier-Stokes equation is
\begin{equation}\label{model_uinst}
du_i = \frac{\partial \lra{ u_iu_k }}{\partial x_k}\, dt
- u_k\frac{\partial \lra{ U_i }}{\partial x_k}\, dt
+ G_{ik}u_k \, dt + A_{ik}u_k \, dt + \sqrt{C_0\lra{ \epsilon }}dW_i.
\end{equation}
The first two terms on the rhs are exact and are independent of the form
of the stochastic model.    The different SDEs are defined in the It\^o sense and the
derivatives of the products $u_iu_j$ are obtained from It\^o's formula
\begin{equation}
d(u_iu_j)=u_idu_j + u_jdu_i + C_0\lra{\epsilon}\, dt \delta_{ij}.
\end{equation}

The mean second-order equations are then
\begin{equation}
\begin{split}
\frac{\partial \lra{ u_iu_j }}{\partial t}
+ \lra{ U_k }
\frac{\partial \lra{ u_iu_j }}{\partial x_k}
+\frac{\partial \lra{ u_iu_ju_k }}{\partial x_k} =&
-\lra{ u_iu_k }\frac{\partial \lra{ U_j }}
{\partial x_k} - \lra{ u_ju_k }\frac{\partial \lra{ U_i}}
{\partial x_k} \\
& + G_{ik}\lra{ u_ju_k } + G_{jk}\lra{ u_iu_k } \\
& + \nu\frac{\partial \lra{ u_iu_j }}{\partial x_k^2} 
+ C_0\lra{ \epsilon } \delta_{ij}.
\end{split}
\end{equation}

\section{Calculus of the stochastic integrals}
Here, it is explained how the stochastic integrals, \textit{appearing
in the analytical solutions of the equation system with constant
coefficients}, can be re-arranged (by stochastic integration by
parts), to yield the covariance matrix.

\subsection{Integration by parts} \label{app:intpart}
Let $X(t)$ and $Y(t)$ be two diffusion processes. One can show that
(see for example Klebaner \cite{Kle_99}), in the It\^o sense,
\begin{equation} \notag
X(t)Y(t)-X(t_0)Y(t_0) = \int _{t_0}^{t}X(s)\,dY(s) + \int
_{t_0}^{t}Y(s)\,dX(s) + [X,Y](t),
\end{equation}
where $[X,Y](t)$ is the quadratic covariation of $X(t)$ and $Y(t)$ on 
$[t_0,t]$. In the case where one of the processes is deterministic,
$[X,Y](t)=0$. In the frame of our study, where the integrated
variable is always a deterministic function, one can therefore apply
integration by parts as in classical differential calculus.

In fact, in the analytical solutions of the equation system with
constant coefficients, one encounters multiple stochastic integrals of
the type
\begin{equation} \label{eq:int_mul}
I=\int_{t_0}^{t} \exp(-s/a)\left( \int_{t_0}^{s} \exp(s'/b)\, dW(s')
\right) \, ds,
\end{equation}
where $(a,b) \in \mathbb{R}^{+2}$. By setting
\begin{align}
& F(s) = \int_{t_0}^{s} \exp(s'/b)\, dW(s')
         \,\Longrightarrow\, dF(s)=\exp(s/b)\,dW(s),  \notag \\
& dG(s) = \exp(-s/a) \,\Longrightarrow
         \, G(s) = -a\, \exp(-s/a)\,ds, \notag
\end{align}
and applying integration by parts, one obtains
\begin{equation} \label{eq:int_sim}
I= - a\, \exp(-t/a)\int_{t_0}^{t} \exp(s/b)\, dW(s) + a \int_{t_0}^{t}
\exp(-s/a)\, \exp(s/b)\, dW(s).
\end{equation}
Therefore, by stochastic integration by parts, the multiple integral
given by Eq. (\ref{eq:int_mul}) can be written as the sum of two
simple stochastic integrals, Eq. (\ref{eq:int_sim}).

\subsection{Derivation of the covariance matrix} \label{app:covmat}
By using the results of the previous subsection and the main
properties of the It\^o integral, that is, linearity, the zero mean
property,
\begin{equation} \notag
\lra{\int_{t_0}^{t} X(s)\,dW(s)} = 0,
\end{equation}
and the isometry property,
\begin{equation} \notag
\lra{\int_{t_0}^{t_1} X(s)\,dW(s)\int_{t_2}^{t_3} Y(s)\,dW(s)} =
\int_{t_2}^{t_1} \lra{X(s)Y(s)}\, ds,
\end{equation}
with $t_0 < t_2 < t_1 < t_3$. From the zero mean
property, it follows that the first order moments are equal to
zero. For the second order moments (covariance matrix), the previous
properties give the following equality
\begin{equation} \label{eq:res2}
\begin{split}
\lra{ \left( \sum_m g_{m}(t) \int _{t_0}^{t} f_{m}(s) \,dW(s)
\right)^2} & = \\ \sum_m g^2_{m}(t) & \int_{t_0}^{t} f^2_{m}(s) \,ds
+ 2 \sum _{m<k} g_{m}(t)\,g_{k}(t)\int_{t_0}^{t}f_{m}(s)\,f_{k}(s)\,ds.
\end{split}
\end{equation}
where $g_{im}(t)$ and $f_{im}(t)$ are deterministic functions of
time. 

\section{Simulation of a Gaussian vector} \label{app:choleski}
Let $\mb{X}=(X_1,\dots,X_d)$ be a Gaussian vector defined by a zero
mean and a covariance matrix $C_{ij}=\lra{X_i X_j}$. For all positive
symmetric matrix (such as $C_{ij}$), there exists a (low or high)
triangular matrix $P_{ij}$ which verifies
\begin{equation} \notag
\mb{C}=\mb{P}\mb{P}^t \,\Longrightarrow\, C_{ij} = \sum _{k=1}^{d}
P_{ik}P_{jk}.
\end{equation}
$\mb{P}$ is given by the Choleski algorithm (here for the low
triangular matrix)
\begin{equation}
\begin{split}
& P_{i1} = \frac{C_{i1}}{\sqrt (C_{11})},
           \quad 1 \leqslant i \leqslant d \notag \\
& P_{ii} = \left(C_{ii}-\sum _{j=1}^{i-1}P_{ij}\right)^{1/2},
           \quad 1 < i \leqslant d \\
& P_{ij} = \frac{1}{P_{jj}}
           \left(C_{ij}-\sum _{k=1}^{j-1}P_{ik}P_{jk}\right),
           \quad 1 < j < i \leqslant d \notag \\
& P_{ij} = 0, \quad i < j \leqslant d \notag .
\end{split}
\end{equation}
Let $\mb{G}=(G_1,\dots,G_d)$ be a vector composed of independent
$\mc{N}(0,1)$ Gaussian random variables, then it can be shown that the
vector $\mb{Y}=\mb{P}\mb{G}$ is a Gaussian vector of zero mean and
whose covariance matrix is $\mb{C}=\mb{P}\mb{P}^t$. Therefore,
$\mb{X}$ and $\mb{Y}$ are identical, that is,
\begin{equation} \label{eq:simu}
\mb{X}=\mb{P}\mb{G}\,\Longrightarrow\, X_i
      = \sum _{k=1}^{d} P_{ik}G_{k}.
\end{equation}
Eq. (\ref{eq:simu}) shows how the stochastic integrals, obtained in the 
analytical solutions of the system with constant coefficients, can be
simulated.

\bibliographystyle{unsrt}
\bibliography{biblio}

\end{document}